\documentclass[fleqn,usenatbib]{mnras}

\usepackage[T1]{fontenc}
\usepackage{ae,aecompl}

\usepackage{graphicx,color}	
\usepackage[dvipsnames]{xcolor}
\usepackage{amsmath}	
\usepackage{amssymb}	
\usepackage{times}
\usepackage{threeparttable}
\usepackage{epstopdf}
\usepackage{url}
\usepackage{bm}
\usepackage{hyperref}

\newcommand{\eqb}{\begin{eqnarray}}
\newcommand{\eqe}{\end{eqnarray}}

\newcommand{\Lg}{L_\gamma}
\newcommand{\Le}{L_{\rm e}}

\newcommand{\tbmax}{t_{\rm b, max}}
\newcommand{\temin}{t_{\rm e, min}}
\newcommand{\tb}{t_{\rm b}}
\newcommand{\te}{t_{\rm e}}
\newcommand{\tg}{t_{\gamma}}

\newcommand{\etag}{\eta_\gamma}
\newcommand{\fcor}{f_{\rm cor}}
\newcommand{\swift}{\emph{Swift}}

\title[Central engines of collapsar GRBs]{Deciphering the properties of the central engine in  GRB collapsars}

\author[Petropoulou et al.]{
M. Petropoulou,$^{1}$\thanks{E-mail: m.petropoulou@astro.princeton.edu}
P. Beniamini,$^{2}$
G. Vasilopoulos,$^3$
D. Giannios,$^{4}$
R. Barniol Duran$^5$
\\
$^{1}$Department of Astrophysical Sciences, Princeton University, 4 Ivy Lane, Princeton, NJ 08544, USA\\
$^{2}$Division of Physics, Mathematics and Astronomy, California Institute of Technology, Pasadena, CA 91125, USA\\
$^3$Department of Astronomy, Yale University, PO Box 208101, New Haven, CT 06520-8101, USA\\
$^4$ Department of Physics, Purdue University, 525 Northwestern Avenue, West Lafayette, IN, 47907, USA \\
$^{5}$Department of Physics and Astronomy, California State University, Sacramento, 6000 J Street, Sacramento, CA 95819, USA }

\date{Accepted XXX. Received YYY; in original form ZZZ}

\pubyear{2020}

\begin{document}
\label{firstpage}
\pagerange{\pageref{firstpage}--\pageref{lastpage}}
\maketitle

\begin{abstract}
The central engine in long gamma-ray bursts (GRBs) is thought to be a compact object produced by the core collapse of massive stars, but its exact nature (black hole or millisecond magnetar) is still debatable. Although the central engine of GRB collapsars is hidden to direct observation, its properties may be imprinted on the accompanying electromagnetic signals. We aim to decipher the generic properties of central engines that are consistent with prompt observations of long GRBs  detected by the Burst Alert Telescope (BAT) on board the \emph{Neil Gehrels Swift Observatory}. Adopting a generic model for the central engine, in which the engine power and activity timescale are independent of each other, we perform Monte Carlo simulations of long GRBs produced by jets that successfully breakout from the star. Our simulations consider the dependence of the jet breakout timescale on the engine luminosity and the effects of the detector's flux threshold. The two-dimensional (2D) distribution of simulated detectable bursts in the gamma-ray luminosity versus gamma-ray duration plane  is consistent with the observed one for a range of parameter values describing the central engine. The intrinsic 2D distribution of simulated collapsar GRBs peaks at lower gamma-ray luminosities and longer durations than the observed one, a prediction that can be tested in the future with more sensitive detectors. Black-hole accretors, whose power and activity time are set by the large-scale magnetic flux through the progenitor star and stellar structure, respectively, are compatible with the properties of the central engine inferred by our model.  
\end{abstract}


\section{Introduction}
The connection between long-duration gamma-ray bursts (GRBs) and the core collapse of massive stars is well established \citep{Woosley93,Stanek+03,Hjorth+03,Woosley&Bloom06}. The collapse results in the formation of a compact star, either a black hole or a rapidly rotating strongly magnetized neutron star (a millisecond magnetar), powering the relativistic jet that produces the GRB emission. The nature of the compact object, known as the GRB central engine, has been a topic of great debate since more than a quarter of a century ago \citep{Paczynski1991,Usov1992,Kluzniak1998}.

As the central engines of GRBs are not directly observable, there have been many attempts to characterize the nature of the central engine based on indirect evidence, such as the appearance of X-ray flares and plateaus in the late-time GRB afterglow light curves (e.g., \citealp{King2005,Dai2006,Perna2006,Proga2006,Zhang2006,Liang2006,Troja2007}). Assuming that these features are indicative of late-time activity of the central engine, inferences about the nature of the latter can be made \citep[e.g.,][]{2014ApJ...785...74L, 2014ApJ...787...66Z, Li2018}. It is still, however, possible to naturally account for the same features without invoking late-time central engine activity \citep{BK2016,BM2017,BDDM2020}. 

Many predictions of the black-hole or magnetar scenarios for the properties of the GRB prompt emission are model-dependent and require a description of the jet energy dissipation and the emission mechanisms, both of which are not fully understood at this point (see \citealt{kumarandzhang2015} for a recent review). The two central engines do, however, differ in certain generic aspects. First, the black-hole and magnetar models differ in what keeps the engine going which, in the first case, is mass accretion onto the compact object and, in the second case, is the compact object's fast rotation. 
Hybrid cases that involve fallback accretion onto a magnetar are also possible, but even then the energy cannot exceed by much the rotational energy reservoir \citep{MBG2018}. In general, black-hole engines are expected to have a wider range of energy reservoirs than magnetars, in which the engine's energy is limited by the initial rotational energy of the magnetar ($\sim 10^{52}$~erg) and the maximum radiated gamma-ray energy is $\lesssim 5\times 10^{51}$~erg \citep{BGM2017}.
Additionally, the relation between the power and the active time of the central engine is expected to differ between black-hole and magnetar scenarios. In the latter scenario, for example, the engine power and activity timescale are related, because both depend on the magnetar's  spin frequency and magnetic field strength. As a result, a tight correlation between the observed gamma-ray luminosity and duration is expected, unless there is a large scatter in the birth properties of magnetars in the long GRB population. If the central engine is a black hole, jets can be powered through neutrino annihilation  \citep[e.g.,][]{Eichler1989, Popham1999, chen2007} or through magnetohydrodynamical (MHD) mechanisms \citep[e.g.,][]{Narayan1992, Meszaros1997}, making different predictions about the jet luminosity and duration and its dependence on physical parameters (e.g., mass accretion rate and stellar structure). Since the neutrino annihilation model appears insufficient in explaining the power of longer bursts \citep{Kawanaka2013, Leng2014}, we will limit our discussion here to MHD models for the jet launching.

In light of the above, it is constructive to consider the generic properties of central engines that are consistent with prompt observations of GRB collapsars. According to the collapsar model for GRBs \citep{macfadyen1999,macfadyen2001}, a jet launched at the core of a collapsing star has to drill its way through the stellar envelope, and break out of the surface before producing the observed gamma-ray signal. Previous works have investigated the imprint of the jet propagation within the collapsing star on the distribution of prompt gamma-ray durations. \cite{bromberg2012,bromberg2013} proposed that the prompt gamma-ray duration distribution of collapsar GRBs exhibits a plateau, extending from the typical jet breakout time $\sim 50$~s, down to much shorter timescales. This idea has been also invoked to understand low-luminosity GRBs as jets that have barely failed to break out \citep{bromberg2011}, to propose a unified picture for low-luminosity and long GRBs \citep{nakar2015}, and to argue that failed jets may operate in all Type Ib/c supernovae \citep{sobacchi2017}. Previous studies assumed a common jet breakout time for all collapsar GRBs, although differences in the properties of their central engines
should yield different breakout times; more powerful jets propagate more easily through the star and break out from it much quicker than weaker jets \citep[e.g.][]{zhang2003, morsony2007, mizuta2009, lazzatietal2012}. Both analytical estimates \citep{bromberg2011, bromberg2011b} and numerical simulations \citep[see][and references
therein]{lazzatietal2012} suggest that the jet breakout time depends upon the  isotropic power of the central engine as $\propto\Le^{-\chi}$, with $\chi \sim 1/3-1/2$ depending on properties of the stellar envelope (e.g., density profile, radius and mass) and/or the properties of the jet (e.g., collimation). \cite{PBDG2017} took into account the luminosity dependence of the breakout time, and demonstrated that the observed broken power-law GRB luminosity function can be the outcome of the jet-envelope interaction for central engines described by a single power-law luminosity distribution. By matching the parameters of the model-predicted GRB luminosity function to the observed one, \cite{PBDG2017} derived a mono-parametric distribution of gamma-ray durations, and inferred the maximum jet breakout time, which was the single tunable parameter of the model. 

Here, we advocate that much more information about the central engine properties can be gleaned by considering the full two-dimensional (2D) distribution of isotropic gamma-ray luminosities $\Lg$ and (rest-frame) durations $\tg$ (see also \citealt{BBPG2020} for applications in the short GRB context). We expand upon the analytical work of \cite{PBDG2017} 
by comparing the 2D distributions (in the $\Lg-\tg$ plane) of simulated and observed long GRBs, detected by the Burst Alert Telescope (BAT) of the \emph{Neil Gehrels Swift Observatory}. To do so, we perform Monte Carlo simulations of long GRBs in the context of a generic central engine model, where the engine power and activity timescale are independent of each other. The Monte Carlo approach also allows us to relax the main simplifying assumptions of the analytical model of \cite{PBDG2017}, namely the universal radiative efficiency among bursts and the completeness of the \swift-BAT sample with respect to the GRB duration. Our simulations yield (for a range of parameter values) 2D distributions that capture the general features of the observed GRB distribution in the $\Lg-\tg$ (e.g., scatter,  correlation, and range of luminosities and durations) and make predictions for those to be detected by future, more sensitive, missions. 

This paper is structured as follows. 
In Section \ref{sec:data} we present the sample of \swift-BAT bursts used in our analysis. In Section \ref{sec:model} we briefly describe the model of the GRB central engine and continue in Section \ref{sec:MC} with a description of our simulations and methodology. In Section \ref{sec:results} we present the results of our simulations. In Section \ref{sec:discuss} we present possible caveats in our analysis and discuss the predictions and implications of our simulations. We conclude in Section~\ref{sec:conclusions} with a summary of our work.

\section{Sample}\label{sec:data}
We use publicly available data from the GRB  archive\footnote{\url{https://swift.gsfc.nasa.gov/archive/grb_table/}} of the \emph{Neil Gehrels Swift Observatory} \citep{Gehrels2004}. We select long GRBs (i.e., bursts with observed $T_{90}\ge2$~s) detected by the \swift \, Burst Alert Telescope (BAT) from 2005 to 2019 with redshift information (either spectroscopic or photometric) and fluence estimation. We also exclude bursts for which only lower limits on $T_{90}$ are available. These cuts result in a sample of 326 bursts (i.e., $\sim27\%$ of \swift-BAT long GRBs). 

To estimate the bolometric isotropic  gamma-ray luminosity, $\Lg$, we use the BAT (energy) fluence $S$ in the 15--150 keV energy range, and the observed burst duration $T_{90}$,
\begin{eqnarray}
\label{eq:Liso}
\Lg(z) =  \frac{4\pi d^2_L(z) S}{T_{90}} \fcor(z),
\end{eqnarray}
where $d_L(z)$ is the luminosity distance\footnote{We adopt a Cosmology of a flat Universe with $H_0 = 69.6$~km s$^{-1}$ Mpc$^{-1}$, $\Omega_{\rm M} =0.31$, and $\Omega_\Lambda =0.69$ \citep{Bennett2014}.} of a burst at redshift $z$, and $\fcor(z)$ is the $k-$correction factor in the rest-frame $1$ keV -- 10 MeV band \citep[e.g.,][]{Bloom2001}. 
From our initial sample of 326 bursts, we select those whose spectrum\footnote{We use spectral fits for photon spectra made with the $T_{100}$ duration, while noting that our main conclusions would not change if spectra from different durations were used.} was fitted either with a power law or a cutoff power law (without checking the quality of the best-fit), and end up with a final sample of 291 long GRBs. For each burst, we compute the correction factor (and estimate the 90\% confidence region) using the best-fit spectral  parameters (and 90\% uncertainties), as reported in the Third \swift-BAT GRB Catalog\footnote{\url{https://swift.gsfc.nasa.gov/results/batgrbcat/index_tables.html}} \citep{3rdBAT}. To compute the uncertainty in $\Lg$, we propagate the errors in fluence and $\fcor$ ($T_{90}$ values from the \swift \, online archive are reported without uncertainties). For plotting purposes, we also compute the isotropic burst energy, $E_{\gamma} = \Lg T_{90}/(1+z)$. For most bursts, the errors in luminosity (and energy) are dominated by the uncertainty in the spectral parameters. 

In some of the plots that appear in Section~\ref{sec:results} we also include, for illustration and comparison purposes, 20 short GRBs (sGRBs; $T_{90}<2$ s) with available spectral parameters, measured redshift and fluence ($\sim 1/5$ of the \swift-BAT sGRB sample). We estimate their isotropic gamma-ray luminosity and errors as described above. 

\section{A generic model for the central engine}\label{sec:model}
Jets launched by the central engine in collapsar GRBs have to drill through the collapsing star in order to break out of it and produce the gamma-ray signal while the central engine
is still active. Here, we adopt a generic scenario for the central engine where its power and  activity timescales are independent of each other \citep{bromberg2012, sobacchi2017, PBDG2017}.

Assuming that the jet propagation time to the gamma-ray production site is negligible\footnote{This is also supported by the fact that the duration of a single GRB pulse is typically much smaller than the inferred breakout time.}, the rest-frame duration of the prompt gamma-ray emission is given by $\tg=\te-\tb$, where $\te$ and $\tb$ are the engine activity and jet breakout times, respectively. For $\te < \tb$, the jet fails to break out from the star and produce a typical GRB (i.e., failed jet). Relativistic hydrodynamic simulations of jet propagation in collapsars have shown that more powerful jets can break out of the stellar envelope more easily than weaker jets. Hence, we use the terms jet luminosity and engine power interchangeably. The breakout time $\tb$ can be related to the isotropic-equivalent jet luminosity, $\Le$,  as \citep[e.g.,][]{bromberg2011, lazzatietal2012, nakar2015}
 \eqb
 \label{eq:tb}
 \tb=t_0 \left(\frac{\Le}{L_{\rm e,0}} \right)^{-\chi}.
 \eqe 
 where $1/3 \lesssim \chi \lesssim 1/2$, $L_{\rm e,0}$
 is a normalization constant, and $t_0$ is a parameter that encodes information about the jet collimation and the properties of the stellar envelope \citep[e.g.,][]{bromberg2011, bromberg2012}. Because the breakout time is shorter for more powerful engines, the jet-collapsar interaction acts as a filter of less luminous jets and of engines of shorter duration. 
 
 \cite{PBDG2017} argued that the observed broken power-law GRB luminosity function is a natural outcome of this filtering process, and that the shape of the GRB duration distribution can be uniquely determined by the GRB luminosity function. Following \cite{PBDG2017}, we adopt a universal $t_0$ for all GRB collapsars (we discuss the case of a non-universal $t_0$ in Section~\ref{sec:t0}), and consider that the  isotropic engine power follows a power-law distribution between $L_{\rm e,\min}$ and $L_{\rm e,\max}$,
\eqb
\label{eq:Le}
f_{\Le}(\Le)= C_{\Le}\left(\frac{L_{\rm e}}{L_{\rm e, \min}}\right)^{-\alpha},
\eqe 
where $C_{\Le}$ is a normalization constant found by the condition $\int f_{\Le} {\rm d}\Le  =1$. We also consider a power-law distribution of engine durations between $t_{\rm e,\min}$ and $t_{\rm e,\max}$, 
\eqb
\label{eq:te}
f_{\te}(\te)= C_{\te} \left(\frac{\te}{\temin}\right)^{-\beta},
\eqe 
where $C_{\te}$ is a normalization constant that ensures $\int f_{\te} {\rm d}\te =1$. The minimum engine activity time can also be expressed as 
\eqb
\temin & = & t_0 \left(\frac{L_{\rm e,*}}{L_{\rm e,0}}\right)^{-\chi},
\label{eq:temin}
\eqe 
where $L_{\rm e,*}$ is a characteristic luminosity above which all engines produce successful jets, and translates to a break in the GRB luminosity function, as demonstrated in \cite{PBDG2017}. These authors estimated $L_{\rm e,*}=3\times10^{53}$~erg s$^{-1}$ by comparing the results of their empirical engine model to the \swift-BAT duration distribution of collapsar GRBs. 

The adopted relation between $\tb$ and $\Le$ implies that the distribution of breakout times is also a power law, 
\eqb
\label{eq:ftb}
f_{\tb}(\tb)= \frac{\alpha-1}{\chi t_0}\left(\frac{L_{\rm e,0}}{L_{\rm e,min}}\right)^{-\alpha+1} \left(\frac{\tb}{t_0}\right)^{\frac{\alpha-1-\chi}{\chi}},
\eqe 
which is truncated at a maximum breakout time $\tbmax \propto L_{\rm e, min}^{-\chi}$. 

\section{Monte Carlo simulations}\label{sec:MC}
In this section, we describe the Monte Carlo simulations used for the generation of long GRBs according to our generic model for the central engine. We also present the Monte Carlo scheme used to explore the multi-dimensional parameter space of the problem.
\subsection{Simulating long GRBs}\label{sec:MC-grb}
For a given set of parameter values describing the GRB central engine, we perform Monte Carlo simulations to determine the isotropic gamma-ray luminosities and gamma-ray durations of successful long GRBs, using the following procedure:
\begin{enumerate}
    \item We generate a pair of random numbers according to the engine power and engine activity  time distributions (see equations \ref{eq:Le} and \ref{eq:te}). 
    \item We compute the breakout time using equation \ref{eq:tb}. 
    \item We compute the rest-frame GRB duration as $\tg=\te-\tb$. If $\tg>0$ (i.e., the jet is successful in breaking out of the star), we continue to the next step. Otherwise, we record the simulated burst as failed, we return to step (i) and repeat the process till we simulate a high number of successful bursts (e.g., $N_{\rm s}\gg10^3$). For certain parameter sets, the success rate is practically zero (e.g., less than one per million), which makes this step of the algorithm computationally expensive.
    \item We generate a random number for the gamma-ray efficiency, which we define as $\etag=\Lg/\Le$, from a uniform distribution (in log space) ranging between $\eta_{\gamma,\min}$ and $\eta_{\gamma,\max}$. We discuss the effects of a unique $\etag$ value on our results in Section~\ref{sec:eta}. 
    \item  We compute the  isotropic bolometric gamma-ray luminosity, as $\Lg=\etag \Le$.
    \item We place each successful simulated burst to a redshift $z$. To do so, we generate $N_{\rm s}$ redshifts according to the differential comoving rate of collapsar GRBs at redshift $z$. Henceforth, we adopt the rate of  \cite{wanderman_piran2010} as defined by their equation~2, with parameter values listed in the first column of their Table~1. 
\end{enumerate}
Finally, we define a simulated GRB as ``detectable'', if its gamma-ray flux exceeds a certain threshold, namely $\Lg/4\pi d^2_L(z) \ge F_{\rm lim}(z)\fcor(z)$. The correction factor $\fcor(z)$ is computed assuming a power-law spectrum with photon index $a$ ($N(E)\propto E^a$). The latter is drawn from a normal distribution of random numbers with mean $-1.5$ and standard deviation $-0.6$, similar to the distribution of photon indices found for \swift-BAT bursts when fitted with a single power law \citep{3rdBAT}. To mimic the effects of the detector's flux threshold, we adopt the limiting flux $F_{\rm lim}$ in the 15-150 keV observer band 
\eqb 
F_{\rm lim}(z)= F_0 \left(\frac{10^6 \ {\rm s}} {\tg(1+z)}\right)^{1/2},
\label{eq:minflux}
\eqe 
where $F_0=2.86\times10^{-11}$~erg cm$^{-2}$ s$^{-1}$ is the flux threshold for an exposure time of 1~Ms \citep[see equation 4 in][]{3rdBAT}. Here, we have implicitly assumed that the exposure time is equal to the observed GRB duration, $T_{90}=\tg(1+z)$. Due to the complexity of the BAT trigger algorithm, the dependence of the detector's sensitivity on the photon incidence angle, and the temporal decay of the burst's flux during the exposure time, the minimum detectable flux given by equation (\ref{eq:minflux}) should be treated as a proxy of the true detector's flux threshold.

\begin{table}
    \centering
    \caption{Parameter values estimated by \citet{PBDG2017} for $\chi=1/3$, $L_{\rm e,*}=3\times10^{53}$~erg s$^{-1}$, $L_{\rm e,0}=10^{51}$~erg s$^{-1}$, and fixed radiative efficiency  $\eta_{\gamma}=0.1$.}
    \begin{tabular}{c c}
    \hline
         Parameter  & Value \\
         \hline 
         $\alpha$ & 2.4 \\
         $\beta$  & 4.6 \\ 
         $t_0$  [s] & $150$\\
         $L_{\rm e, \min}$ [erg s$^{-1}$] & $10^{52}$ \\
         \hline 
    \end{tabular}
    \label{tab:P17}
\end{table}

\begin{figure*}
    \centering
    \resizebox{\hsize}{!}{\includegraphics{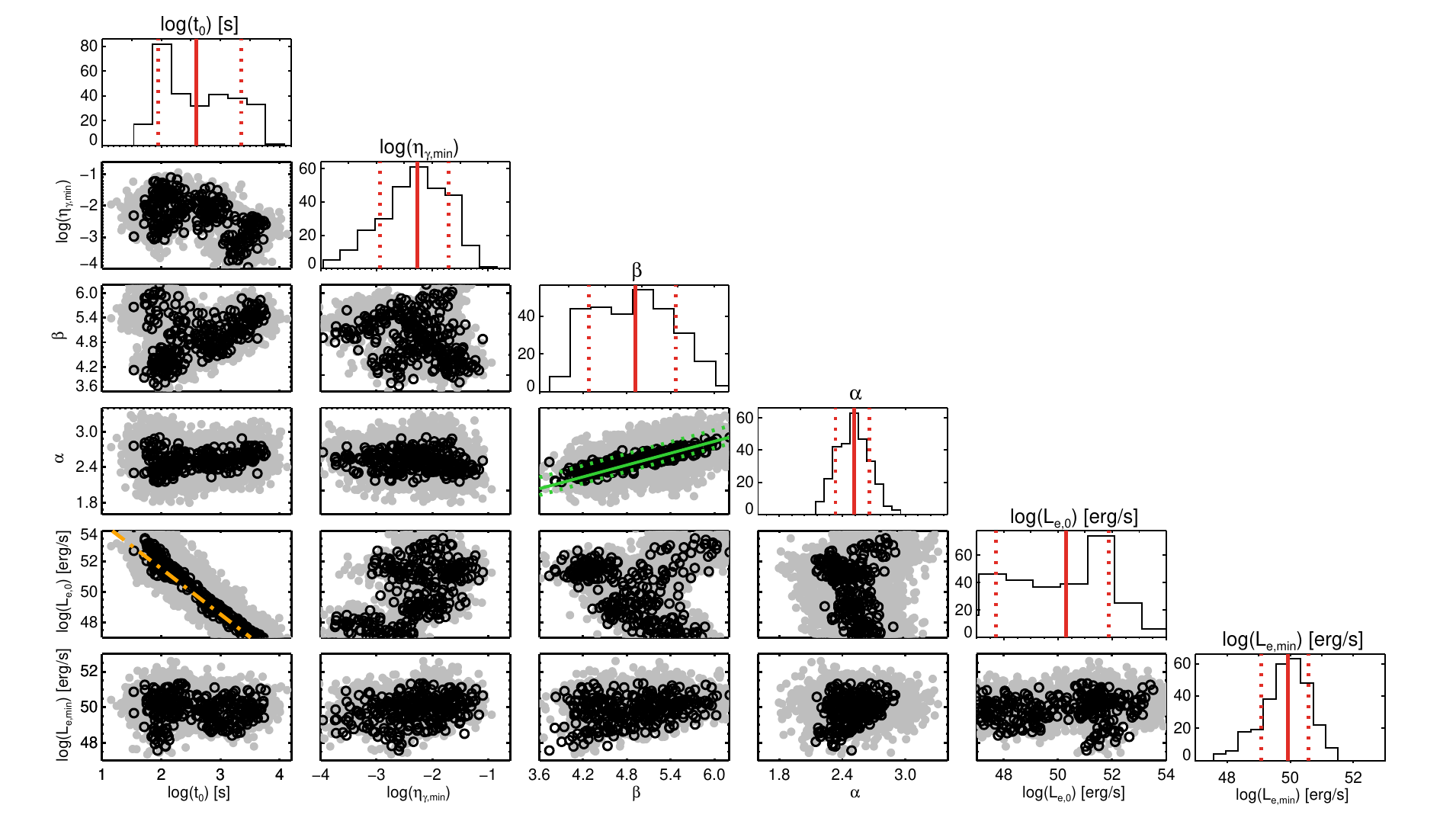}}
    \vspace{-0.5cm}
    \caption{Corner plot summary of parameter values from the Monte Carlo scheme used to compare the $\Lg-\tg$ distributions of the \swift-BAT long GRBs and the simulated detectable bursts. Off-diagonal panels show the 2D distributions for each pair of parameters, with grey (black) color indicating parameter combinations for which we can (cannot) exclude the null hypothesis that the two samples originate from the same population at $>99\%$ confidence. The dash-dotted orange line  shows the expected relation $L_{\rm e,0} = 10^{51}~{\rm erg \, s^{-1}} (150~{\rm s} /t_0)^{-1/\chi}$. For illustration purposes, we also show the relation $\alpha=\beta \chi + (1-\chi) + \alpha_{\rm L}$ for the best-fit value $\alpha_{\rm L}=0.17^{+0.19}_{-0.10}$ from \citet{wanderman_piran2010} (green solid and dashed lines). Panels on the diagonal show the one-dimensional histogram for each model parameter. Each histogram is computed using the black points shown in the off-diagonal plots. Vertical solid and dashed red lines indicate the median and 68\% interval of the distributions, respectively.
    }
    \label{fig:mcmc}
\end{figure*}

\subsection{Parameter exploration}\label{sec:MC-param}
In Table~\ref{tab:P17} we list the parameter values estimated by \cite{PBDG2017} from the comparison of the analytical model to the duration distribution of 319 \swift-BAT long GRBs with redshift information. In this work, we compare our model to the 2D distribution of \swift-BAT long GRBs in the $\Lg-\tg$ plane, we relax the simplifying assumption about universal radiative efficiency among bursts, and perform an exploration of the parameter space using Monte Carlo techniques. Our goal is to search for combinations of model parameters leading to 2D distributions that capture the general features of the observed GRB distribution in the $\Lg-\tg$ plane (e.g., scatter, correlation, and range of luminosities and durations).

Given the small range of theoretically motivated values for $\chi$ and its small effect on the inferred model parameters \citep[see, e.g., Figure 2 in][]{PBDG2017}, we adopt $\chi=1/3$ as a representative value. Additionally, we fix the maximum engine power and engine activity timescale to large enough values, so that the respective $\Lg$ and $\tg$ values of simulated bursts exceed the observed maximum values. We set $L_{\rm e,max}=10^{55}$~erg s$^{-1}$ and $t_{\rm e, max}=10^4$~s, while noting that our main conclusions do not depend on these specific values, since the respective power-law distributions are expected to be soft (see Table~\ref{tab:P17}). 
For an engine with luminosity $L_{\rm e}$, the breakout time depends on the product of two parameters as $t_0\, L_{\rm e,0}^\chi$ (see equation \ref{eq:tb}). As  a result, only the combination of these two parameters can potentially be constrained by our model-to-data comparison. For instance, \cite{PBDG2017} estimated $t_0\sim 150$~s (see Table~\ref{tab:P17}) assuming $L_{\rm e,0}=10^{51}$~erg s$^{-1}$. Here, we treat  both $t_0$ and $L_{\rm e,0}$ as free parameters, and let the latter vary in a reasonably wide range, i.e., from $\sim 10^{47}$ to $\sim 10^{54}$~erg s$^{-1}$. 
Furthermore, the minimum engine timescale depends on the combination of $t_0, L_{\rm e,0}$, and $L_{\rm e,*}$ (see equation \ref{eq:temin}). By letting all three constants vary arbitrarily, we would only increase the number of correlated parameters of the problem without gaining more physical insight. We therefore choose to fix $L_{\rm e,*}$ to the value estimated analytically by \cite{PBDG2017} (see Table~\ref{tab:P17}). Finally, we set $\eta_{\gamma, \max}=0.25$. This upper cutoff is consistent with the distribution of prompt efficiencies inferred from studies of afterglow energetics of \emph{Fermi}-LAT detected GRBs \citep[][]{Beniamini2015, Beniamini2016}.

We then explore a six-dimensional space composed of the following parameters: the power-law indices of the engine power and engine activity time distributions ($\alpha$ and  $\beta$, respectively), the minimum isotropic engine power and minimum radiative efficiency ($L_{\rm e,min}$ and $\eta_{\gamma,\min}$, respectively), and the characteristic breakout timescale $t_0$ of an engine with isotropic luminosity $L_{\rm e,0}$ (see equation \ref{eq:tb}). 
\citet{PBDG2017} showed that the power-law indices $\alpha$ and $\beta$ are not totally unconstrained parameters, as (for fixed $\chi$) they are related to the power-law indices above ($\beta_{\rm L}$) and below ($\alpha_{\rm L}$) the break of the observed luminosity function of collapsar GRBs,  
\eqb 
\label{eq:a}
\alpha & = &\beta_{\rm L} + 1  \\ 
\chi \beta & = &  (\beta_{\rm L}-\alpha_{\rm L}) + \chi.
\label{eq:b}
\eqe 
We therefore limit our search in a range of $\alpha, \beta$ values that is expected from the $1\sigma$ errors on the power-law indices of the luminosity function \citep{wanderman_piran2010}. We note, however, that the power-law indices of the luminosity function do not enter explicitly in our simulations in contrast to \cite{PBDG2017}.

We do not aim to determine the best-fit parameter values of the model, but rather perform a model-to-data comparison and
identify regions of the parameter phase space leading to  distributions of bursts on the $\Lg-\tg$ plane that are consistent with the data. To perform the comparison of the model to the data we adopt a two-sample 2D Kolmogorov-Smirnov test \citep{Peacock1983, FF1987, Press88}, a tool used in variety of astrophysical applications \citep[e.g.,][]{Metchev2002, George2008, Ghisellini2008, Harari2009,Rowlinson2014}. What the 2D KS test ultimately probes is whether the data  are distributed in the $\Lg-\tg$ plane in the same proportion as the model. This is done by checking if there is any quadrant in the 2D plane where the fraction of BAT GRBs is significantly larger than the fraction of simulated detectable GRBs. 

To perform the parameter exploration, we used the following simplified Monte Carlo approach: \begin{enumerate}
\item Let $\bm{p}=\left\{ 
\log(t_0), \log(\eta_{\gamma, \min}), \beta,  \alpha, \log(L_{\rm e,0}),\log(L_{\rm e, min}) \right \}$ be the parameter vector. We randomly choose ten initial sets of parameter values close to (but not exactly the same as) those estimated by \cite{PBDG2017} (see Table~\ref{tab:P17}). For each initial point, we perform $i_{\max}=300$~trials, as described below.
\item For each $\bm{p}_i$ with $i=1, ..., i_{\max}$, we perform a Monte Carlo simulation for producing a sample of $\Lg, \tg$ values for detectable GRBs according to our model (for details, see Section~\ref{sec:MC-grb}). \item We perform a 2D KS test \citep{FF1987} between the sample of \swift-BAT GRBs ($N_{\rm obs}$) and the sample of simulated detectable GRBs ($N_{\rm d}$), and record the value of the test-statistic $Z^{(i)}_{\rm n,2D}$, where $n\equiv N_{\rm d} N_{\rm obs} / (N_{\rm obs}+N_{\rm d})$ is the effective sample size used for the test.  Although the test of \cite{FF1987} is much less computationally demanding than the test proposed by \cite{Peacock1983}, it can still be challenging to perform the test when the sample sizes are large, as in our problem. Thus, we select a random sub-sample from the detectable GRBs with fixed size which is large enough to allow a meaningful comparison to the data but is also sufficiently small  as not to stall the computation. Here, we set $N_{\rm d}=10^3$. 
\item Given a point $\bm{p}_i$, we generate a
new trial point $\bm{p}_{i+1}=\bm{p}_i + \Delta\bm{p}$, where $\Delta\bm{p}$ is randomly drawn from a normal distribution with standard deviation  $\bm{\sigma}=\left \{0.2,0.2,0.2, 0.2, 0.5,0.5\right \}$. 
\item If $Z^{(i+1)}_{\rm n,2D}>Z^{(i)}_{\rm n,2D}$, we return to the previous step and generate another $\bm{p}_{i+1}$, till we find $Z^{(i+1)}_{\rm n,2D}<Z^{(i)}_{\rm n,2D}$. This approach was adopted to avoid large deviations to regions of the parameter space that would make our computational scheme inefficient because of the very low number of successful engines. For the same reason, we avoided large jumps between consecutive trial points (see previous point).
\item We repeat steps (ii) - (iv) until we have created a large set of points $\bm{p}_i$,  for $i=1, \dots, i_{\max}$.
\end{enumerate}

After having created a sample of 3000 points, we compute the critical value of the test statistic, $Z_{\rm n,SL}$, that corresponds to a significance level (SL), following  \citet{FF1987}. Here, we adopt a SL of 99\%. For those parameter sets having $Z^{(i)}_{\rm n,2D}>Z_{\rm n,99}$ we can exclude the null hypothesis that the two samples stem from the same population at $>99\%$ confidence. The method described above allows us to explore the parameter space and identify sub-spaces for which there is $<1\%$ probability that the observed GRB sample and the sample of simulated detectable bursts come from the same population. 

The results of our parameter exploration are summarized in Figure~\ref{fig:mcmc}. Grey filled symbols indicate parameter combinations for which we can exclude the null hypothesis that the two samples are drawn from the same distribution at $>99\%$ confidence. For other parameter combinations (shown with black open symbols), we cannot exclude that the observed and simulated samples are drawn from the same parent distribution. We refer to these parameter sets as ``acceptable''. Histograms of the parameter values from the acceptable trials are shown in the diagonal panels of Figure~\ref{fig:mcmc}. The  vertical solid and dashed lines indicate the median and 68\% interval of the distributions, respectively. 

By comparing the distributions of grey and black symbols, we can infer that the best constrained model parameters are $\alpha$ and $L_{\rm e, min}$. The same applies to the product $t_0 L_{\rm e,0}^\chi$, although this is not explicitly shown in the figure. Our results are not surprising, as these parameters affect directly the distributions of the engine luminosity and breakout times (see equations \ref{eq:Le} and \ref{eq:ftb}). More specifically, the power-law index $\alpha$ affects the number of failed jets up to a certain luminosity (i.e., $N_{\rm f}(L_{\rm e}) \propto L_{\rm e}^{-\alpha+1}$) and the shape of the luminosity distribution of successful engines with $L_{\rm e}< L_{\rm e,*}$. Moreover, both $t_0 L_{\rm e,0}^\chi$ and $L_{\rm e, min}$ determine the maximum breakout time, $\tbmax$, which is imprinted on the shape of the duration distribution at $\tg \gg 10$~s \citep[see Figure~2 in][]{PBDG2017}.
Because the breakout timescale of an engine with given luminosity depends on $t_0 L^\chi_{\rm e,0}$, all acceptable trials should have anti-correlated $t_0, L_{\rm e,0}$ values. Indeed, we find that the $t_0$ and $L_{\rm e,0}$ values of the acceptable trials follow the expected relation $L_{\rm e,0} = 10^{51}~{\rm erg \, s^{-1}} (150~{\rm s} /t_0)^{-1/\chi}$ (see dash-dotted orange line). Without loss of generality, one could fix $L_{\rm e,0}=\mathcal{O}(10^{51})$~erg s$^{-1}$ \citep[see e.g.,][]{bromberg2012} and let $t_0$ vary. In this case, the distribution of acceptable $t_0$ values would become narrower. It is interesting to note that $t_0 \gg 20$~s, unless the characteristic engine luminosity becomes extremely high ($\gg 10^{54}$~erg s$^{-1}$). We discuss the implications of the derived $t_0$ values in Section~\ref{sec:t0}. 

According to the analysis of \citet{PBDG2017}, the gamma-ray duration distribution for $\tg \gtrsim \tbmax$ reflects the distribution of engine times, namely $f_{\tg}(\tg) \propto \tg^{-\beta}$. For the range of acceptable $t_0, L_{\rm e,0}$, and $L_{\rm e, \min}$ values, we find that 68\% of the $\tbmax$ values lies between $\sim 250$~s and $760$~s. Given that the (rest-frame) duration distribution of \swift-BAT GRBs does not extend beyond $\sim2\times10^3$~s (only 5 out of 291 bursts have $\tg>200$~s), we lack the dynamic range for constraining $\beta$, as shown in Figure~\ref{fig:mcmc}. We find that the $\alpha, \beta$ values from all cases are correlated as expected (see equations \ref{eq:a} and \ref{eq:b}). Interestingly, all acceptable cases lie within a very narrow stripe (dotted green lines) in the $\alpha$ versus $\beta$ plot. The width of the stripe is solely determined by the $1\sigma$ errors on the power-law index $\alpha_{\rm L}$ of the GRB luminosity function \citep{wanderman_piran2010}, a result that 
was driven by the model-to-data comparison and not imposed on our simulations. Finally, we find that the minimum radiative efficiency for the acceptable cases spans a wide range of values, suggesting that the latter is either a subsidiary parameter of the model or that the data are not enough yet to constrain it.

Based on the setup of our parameter exploration, we cannot formally exclude the existence of parameter combinations other than those shown in Figure~\ref{fig:mcmc} (black symbols) that can also represent the data. However, we do not expect them to be radically different than those derived here, as most of the parameters (e.g., $t_0 L_{\rm e,0}^\chi$, $\alpha$, and  $L_{\rm e,min}$) are directly related to observables, such as the power-law indices of the GRB luminosity function and the shape of the $t_{\gamma}$ distribution, particularly at the long durations \citep[for analytical expressions, see][]{PBDG2017}.

\section{Results}\label{sec:results} 
As a representative example, we present results of a Monte Carlo realization with $N_{\rm s}=10^6$ successful GRB jets, using one set of plausible parameter values (see Table~\ref{tab:param}) drawn from the 68\% interval of acceptable trials (see previous section). For the adopted parameters, 
we find $N_{\rm f}\sim 3000$ failed GRBs for each successful one, while $\sim 6\%$ of the simulated successful bursts are detectable. The fraction of detectable bursts is found to range between $\sim4\%$ and $\sim8\%$ for other parameter values drawn from the allowed parameter space, and the number ratio of successful to failed GRBs is $\sim0.03\%-0.1\%$.

\begin{table}
    \centering
    \caption{Parameter values for a Monte Carlo simulation of long GRBs presented in Section~\ref{sec:results} as an illustrative example.}
    \begin{threeparttable}
    \begin{tabular}{ccc ccc}
    \hline
    $\alpha$ &  $\beta$  &  $t_0$  [s] &  $L_{\rm e,0}$ [erg s$^{-1}$] & $L_{\rm e, \min}$  [erg s$^{-1}$] & $\eta_{\gamma, \min}$     \vspace{0.05in} \\
    2.4 & 4.6 & 118 & $4\times10^{51}$ & $4\times10^{49}$ &  0.014 \\
         \hline 
    \end{tabular}
    \begin{tablenotes}
    \item{Note --} The values of parameters $\alpha$ to $\eta_{\gamma,\min}$ were chosen from the 68\% interval of acceptable trials (see Figure~\ref{fig:mcmc}). Other parameters used (and kept fixed in the parameter exploration) are:  $\chi=1/3$,  $L_{\rm e, max}=10^{55}$~erg s$^{-1}$,  $L_{\rm e, *}=3\times10^{53}$~erg s$^{-1}$,  $t_{\rm e, max}=10^4$~s, and $\eta_{\gamma, \max}=0.25$.
    \end{tablenotes}
  \end{threeparttable}
\label{tab:param}
\end{table} 

\begin{figure*} 
    \centering
    \includegraphics[width=0.8\textwidth]{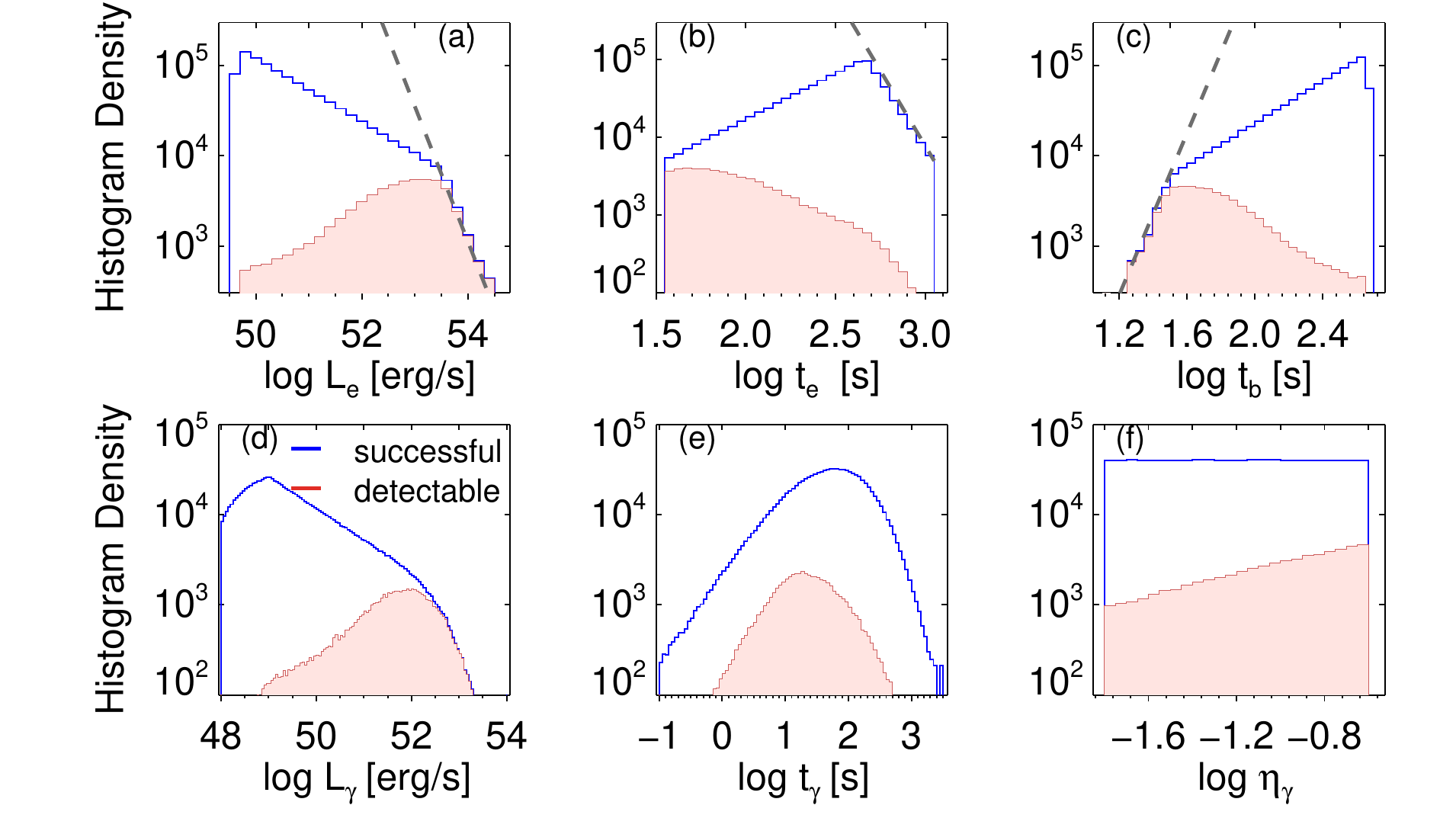}
    \caption{Histograms (in logarithmic scale) of various properties of successful GRBs (blue) and detectable  GRBs (red) from a Monte Carlo simulation with parameters  listed in Table~\ref{tab:param}. From panel (a) to panel (f) we show (in clockwise order) the logarithms of the (isotropic) engine power, engine activity timescale, breakout time, (isotropic) gamma-ray luminosity, gamma-ray duration, and gamma-ray efficiency. Dashed grey lines (top panels) show the intrinsic power-law distributions of simulated engine properties.}
    \label{fig:histo}
\end{figure*}
Figure \ref{fig:histo} shows the histograms of various properties of successful GRBs (blue) and detectable  GRBs (red) from our Monte Carlo simulation. The most powerful engines ($L_{\rm e}> L_{\rm e,*}$) are all successful and  power detectable bursts, whereas the fraction of failed jets increases for $L_{\rm e}< L_{\rm e,*}$ (compare blue histogram and dashed grey line in panel a). Moreover, intrinsically weaker engines power lower luminosity bursts, which are more likely to fall below the flux threshold (compare blue and red histograms in panel a). The distribution of $\te$ for successful engines is also a broken power-law with a break at $\sim10^{2.7}$~s (panel b), which is related to the maximum breakout time of successful engines (panel c). The fraction of successful engines producing detectable GRBs decreases with increasing $\tb$ (panel c), as engines with longer breakout times are weaker and are more likely to power bursts whose flux will be lower than the detector's flux threshold given the redshift evolution of the GRB rate. At high luminosities (i.e., $\Lg \gtrsim 2\times10^{52}$~erg s$^{-1}$), all bursts are detectable, and the distribution of high-luminosity GRBs matches the one for the central engine power  (compare red and blue histograms in panel d). The number of undetectable bursts quickly increases as $\Lg \lesssim 10^{51}$~erg s$^{-1}$, thus changing significantly the shape of the intrinsic luminosity function \citep[see also][]{wanderman_piran2010}. The effects of the flux threshold on the distribution of gamma-ray durations are less pronounced, as  the overall shape of both distributions is similar (compare blue and red histograms in panel e). We note, however, that the average burst duration of detectable bursts is shifted towards shorter timescales compared with the average value for the whole population of successful bursts. Lastly, there is some depletion of detectable bursts towards lower radiative efficiencies, as expected (panel f).

\begin{figure*}
    \centering
    \includegraphics[width=0.47\textwidth]{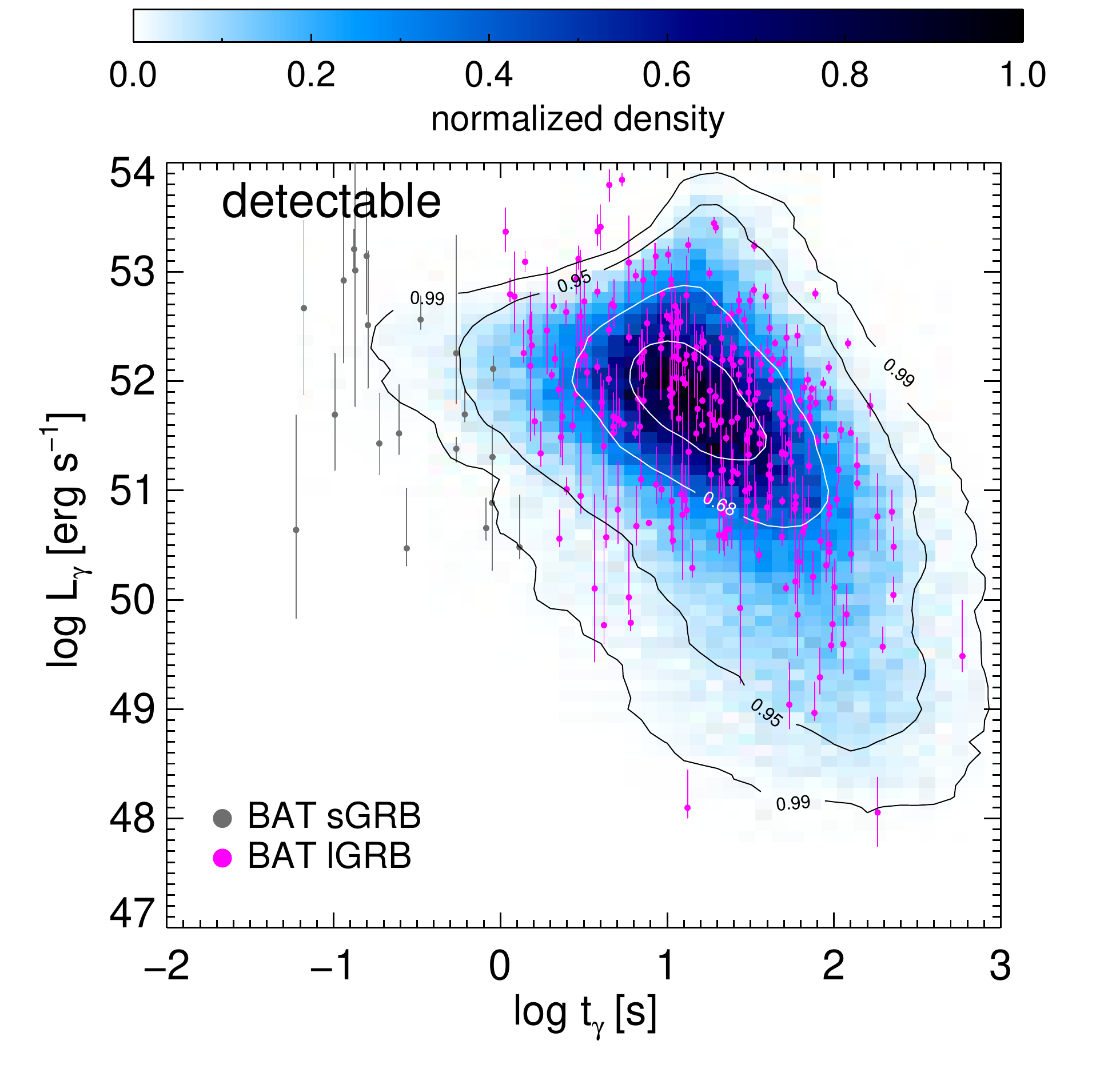}
    \includegraphics[width=0.47\textwidth]{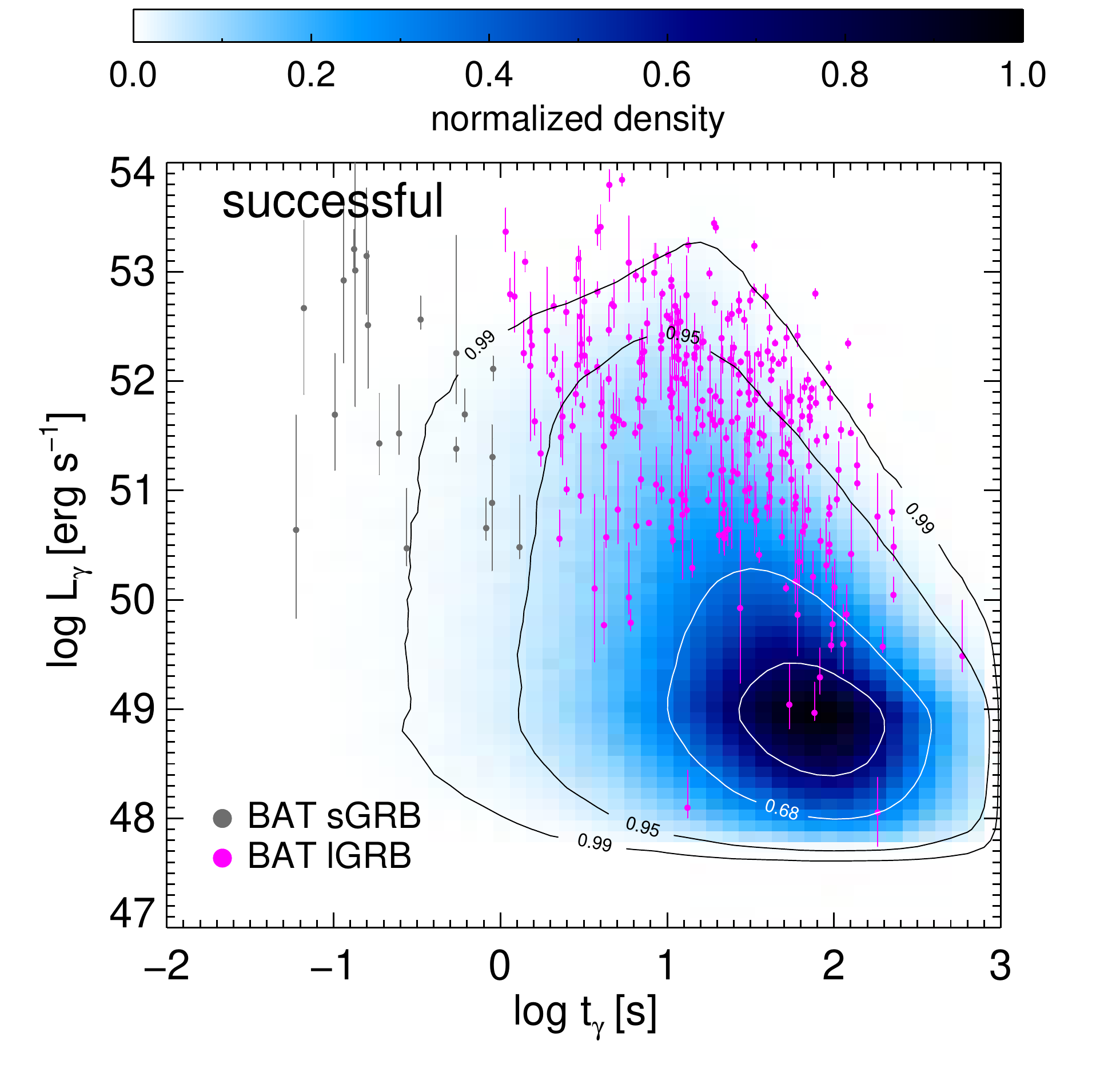}
    \caption{Left panel: Density map (coloured region) and density contours (solid lines) of {\sl detectable} bursts in the $\Lg-\tg$ plane from one Monte Carlo realization (for parameters, see Table~\ref{tab:param}). For comparison, long-duration ($T_{90}\ge 2$~s, magenta symbols) and short-duration ($T_{90}<2$~s, grey symbols) \swift-BAT GRBs with measured redshifts are overplotted. Right panel: Same as in the left panel, but for all simulated {\sl successful} bursts.}
    \label{fig:densmap}
\end{figure*} 

\begin{figure*}
    \centering
    \includegraphics[width=0.47\textwidth]{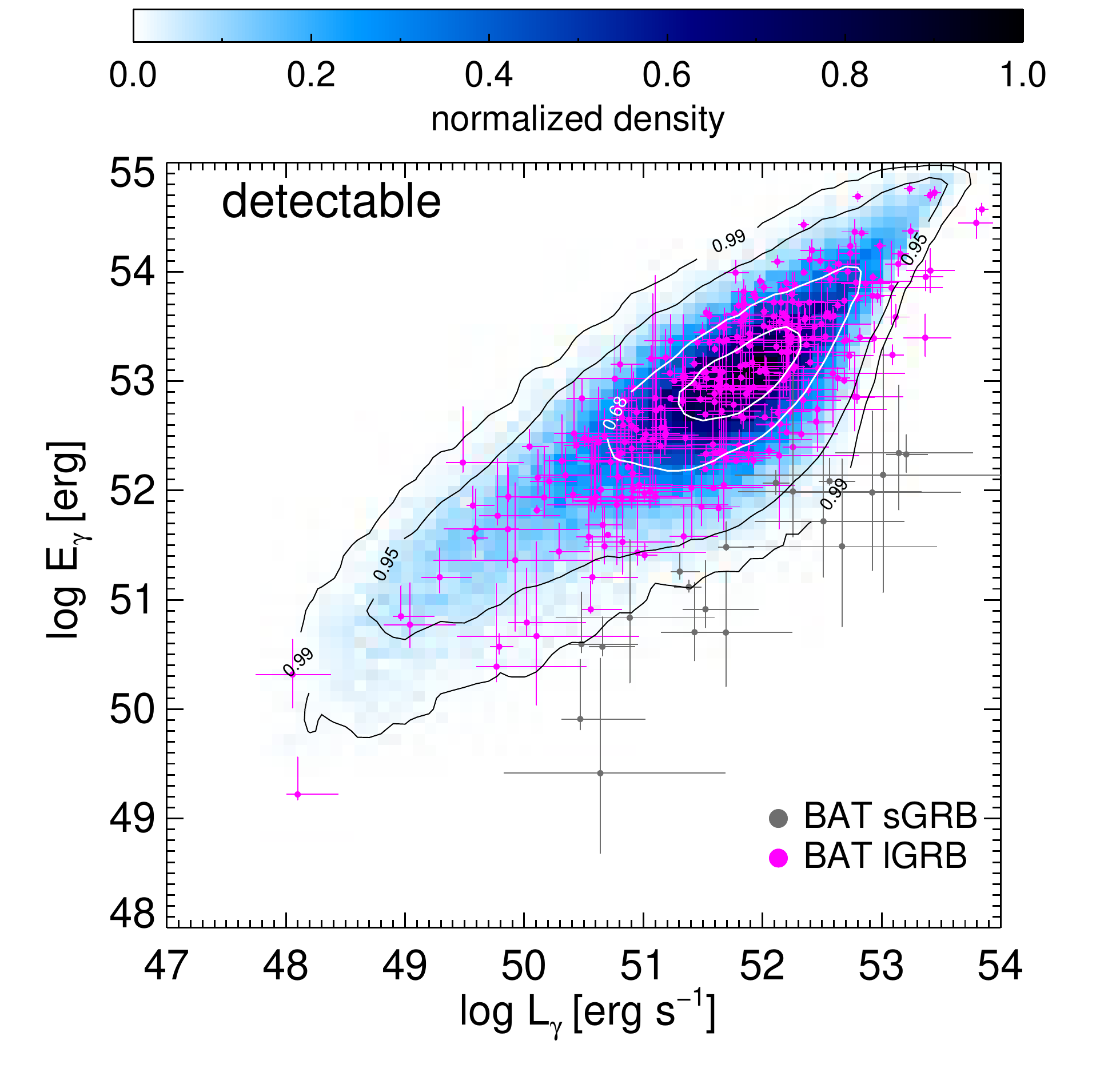}
    \includegraphics[width=0.47\textwidth]{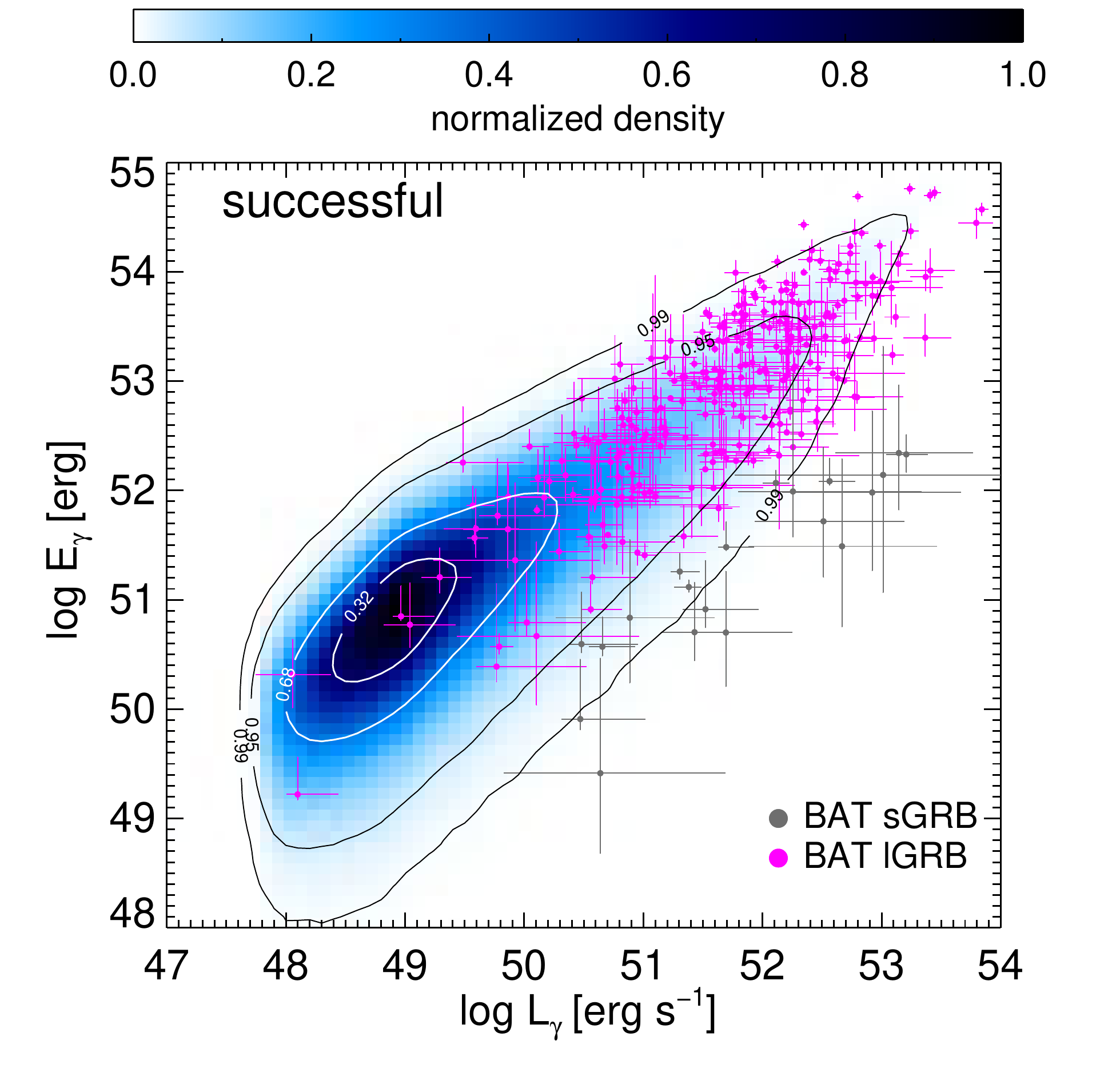}
    \caption{Same as in Figure~\ref{fig:densmap}, but in the $E_{\gamma}-\Lg$ plane.}
    \label{fig:densmap-Eiso}
\end{figure*} 

Figure~\ref{fig:densmap} shows density maps of the simulated detectable bursts (left panel) in the $\Lg-\tg$ plane overlaid with our \swift-BAT sample of long GRBs (magenta symbols). For comparison purposes, we also include 20 sGRBs ($T_{90}<2$ s) with available spectral parameters, measured redshift and fluence (grey symbols). These are also relevant in our discussion about the completeness of our collapsar sample (for details, see Section~\ref{sec:caveats}).

\begin{figure*}
    \centering
    \includegraphics[width=0.33\textwidth]{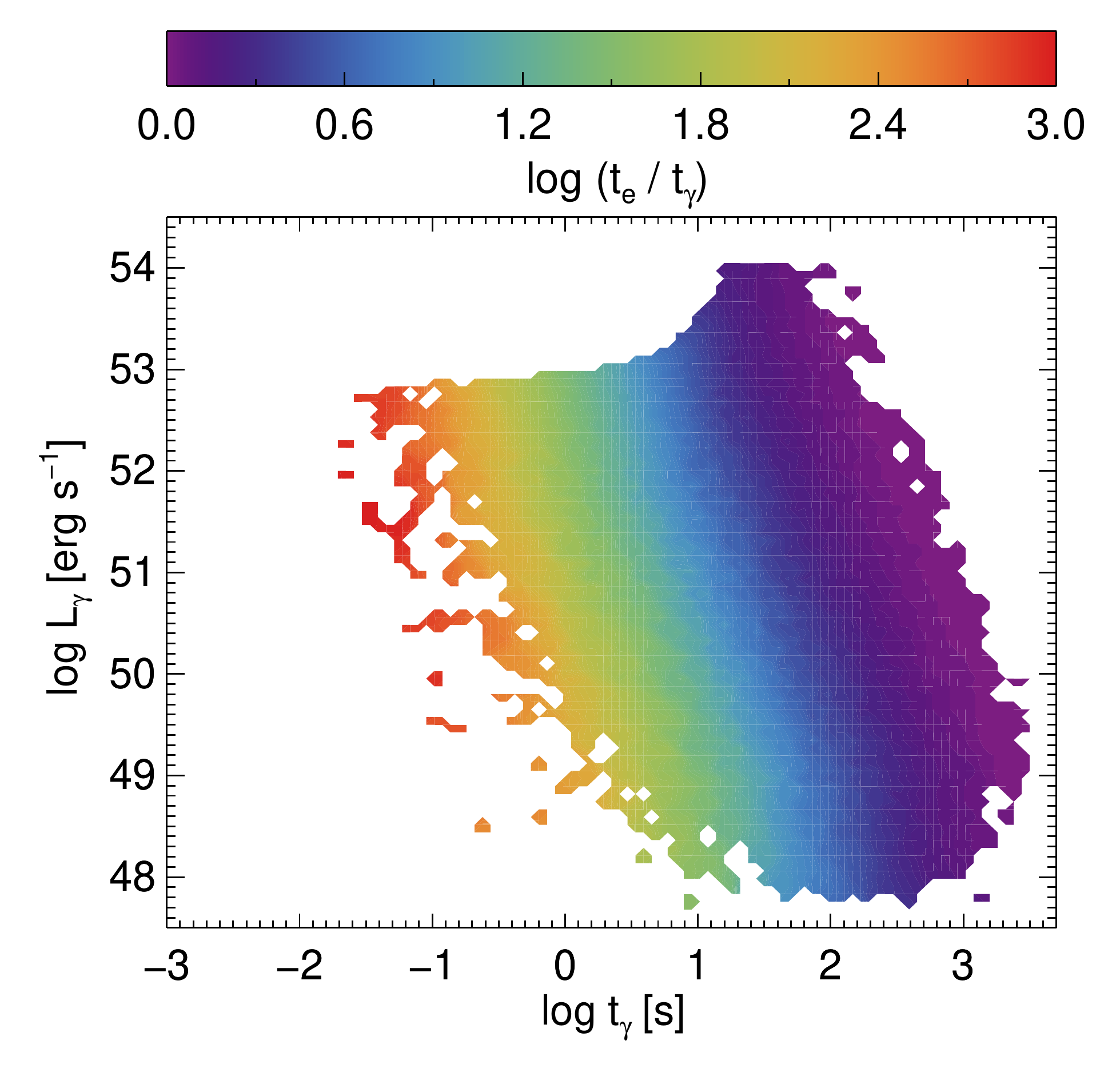}
    \includegraphics[width=0.33\textwidth]{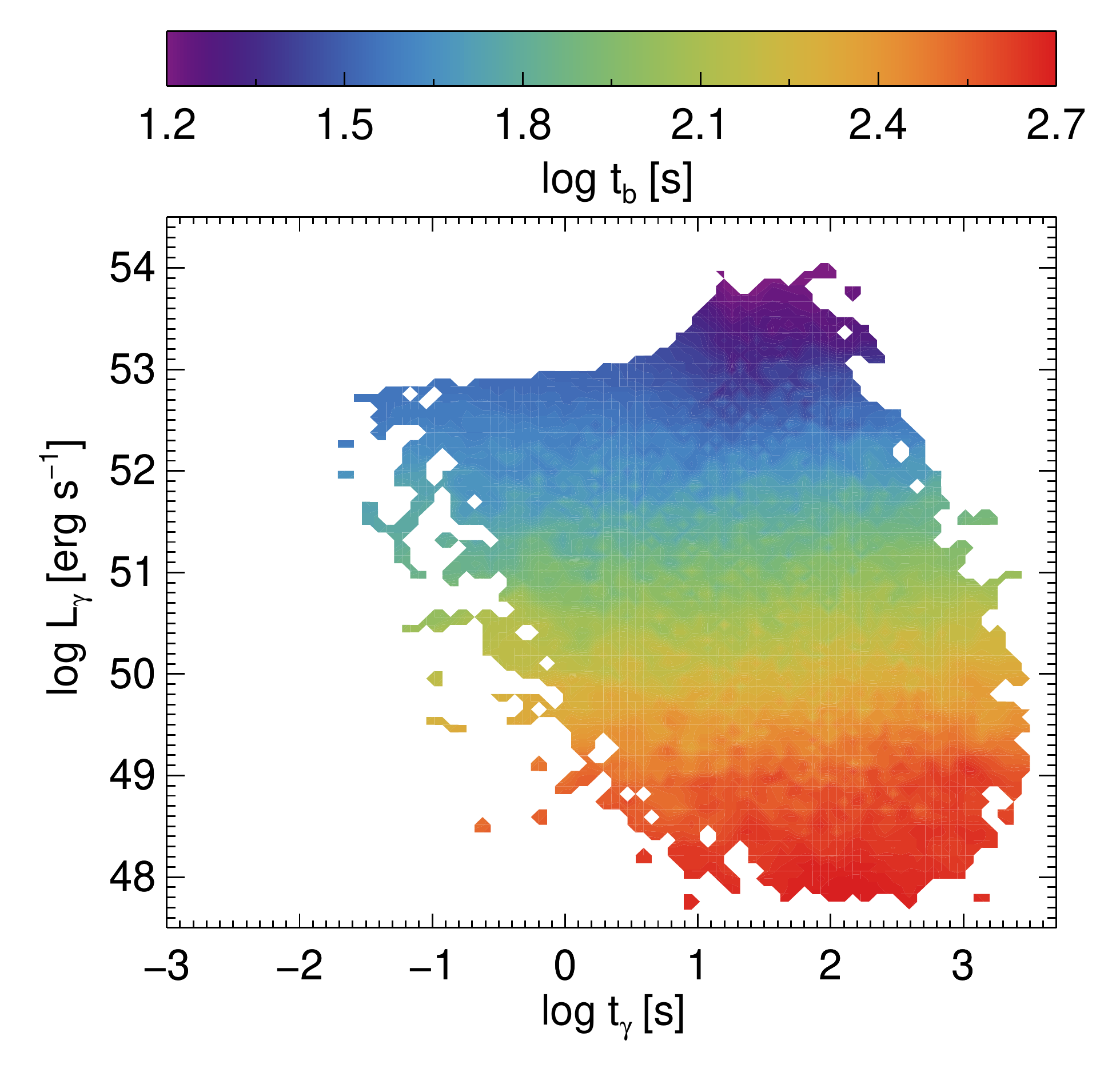}
    \includegraphics[width=0.33\textwidth]{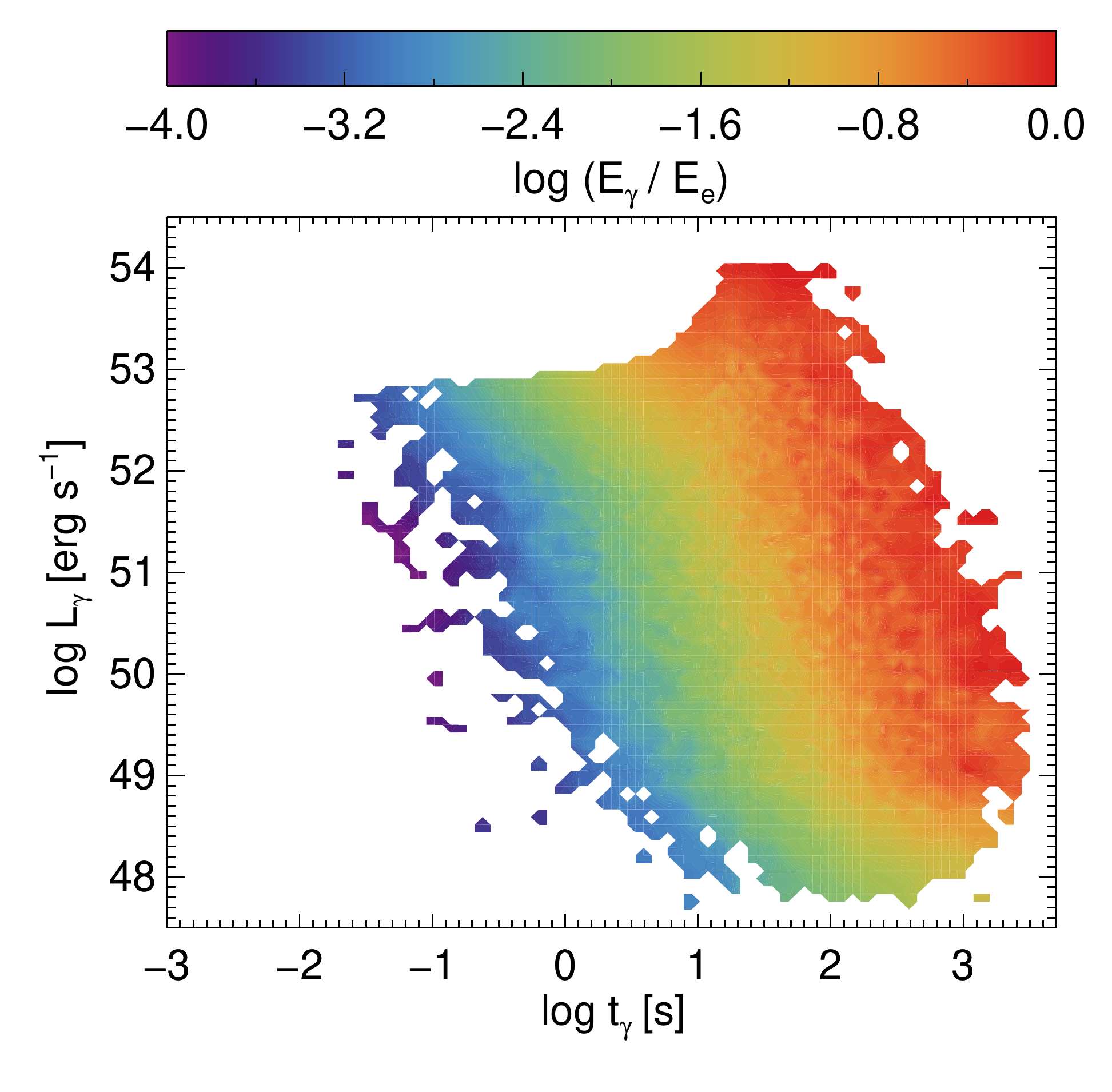}
    \caption{Left panel: Contour plot of the ratio of engine activity timescale and gamma-ray duration (in logarithmic scale) for all detectable bursts from our Monte Carlo simulation. Middle and right panels: Same as in the left panel, but for the breakout time and the ratio $E_{\gamma}/E_{\rm e}$, respectively. Here, $E_{\rm e}$ is the (isotropic equivalent) energy of the central engine.
    }
    \label{fig:heatmap}
\end{figure*}

Our model can reproduce the main features of the observed $\Lg-\tg$ distribution of long GRBs, such as the location of the maximum density and the scatter of the 2D distribution.  To better quantify the model-to-data comparison, we performed a 2D KS test between the sample of 291 long GRBs and $10^4$ equally-sized  sub-samples of detectable simulated bursts that were randomly selected out of a total of $N_{\rm d}\sim 6\times10^4$. We found that we cannot exclude the hypothesis that the samples of simulated and observed GRBs come from the same population (i.e., $Z_{\rm n,2D}<Z_{\rm n,99}$) in 99.84\% of the tests we performed. The median value of the ratio $Z_{\rm n,2D}/Z_{\rm n,99}$ is 0.61, while 68\% of the values lie between 0.52 and 0.72. There is a handful of observed long bursts (6/291) with $\Lg >10^{53}$~erg s$^{-1}$ and $\tg <10$~s, but no simulated bursts are found in this part of the phase pace. This discrepancy is not enough for making us reject the null hypothesis of a 2D KS test (i.e., the samples of observed and detectable GRBs are drawn from the same distribution). Moreover, it can be alleviated by simply considering a slightly higher value of the characteristic engine luminosity (e.g., $L_{\rm e,*}=(6-10)\times 10^{53}$~erg s$^{-1}$). This choice would result in a larger fraction of high-luminosity successful engines, and in turn GRBs, without altering any other major features of the simulated distributions. The model does not predict luminous bursts with long durations (upper right corner) in agreement with the data. The lack of bursts in this part of the phase space can be understood as follows. The most luminous GRBs produced by the most powerful engines ($\Lg = \etag L_{\rm e}$) that are rare (see equation \ref{eq:Le}). At the same time, the respective breakout times are short ($\tb \propto L_{\rm e}^{-\chi}$). As a result, long gamma-ray durations are equivalent to long engine timescales (see also Figure~\ref{fig:heatmap}). Taking into account that the distribution function of engine timescales is also steep (see equation \ref{eq:te}), powerful engines with long activity timescales are very rare in our model, thus explaining the lack of bursts in the upper right corner of the $\Lg-\tg$ plane. 

The intrinsic  2D distribution of bursts predicted by our model is displayed in the right panel of Figure~\ref{fig:densmap}, where we plot the density map of all successful simulated bursts. At high luminosities almost all successful bursts are detectable regardless of their duration. Although this is not evident from this figure because of the adopted color scale, it can be clearly seen by comparing the red and blue histograms in panel d of Figure~\ref{fig:histo}. On the contrary, a high fraction of GRBs at lower luminosities (i.e., $\Lg \lesssim 10^{52}$~erg s$^{-1}$) cannot be detected (see also panel d in Figure~\ref{fig:histo}). As a result, the peak of the intrinsic $\Lg-\tg$ distribution shifts to longer durations and lower luminosities compared with the peak of the 2D distribution of detectable bursts. The observed distribution of \swift-BAT bursts is still the tip of the iceberg, and according to our model we should start detecting more GRBs with $\tg \sim 30-100$~s and $\Lg \sim 10^{49}-10^{50}$~erg s$^{-1}$, as the sensitivity of X-ray satellites improves (see also Section~\ref{sec:sensitivity}). We also note that the flux threshold imposed in our simulations does not have a strong effect on the shape of the $\tg$ distribution, but has a strong impact on the $\Lg$ distribution of detectable bursts below $\sim 3\times10^{52}$~erg s$^{-1}$, as shown also in Figure~\ref{fig:histo}.  

The concave shape of the high-luminosity part of the density map (see left panel in Figure~\ref{fig:densmap}) is an intrinsic feature of the model, which is unrelated to the flux threshold and the $k$-correction used in our simulations. The $k$-correction for simulated bursts is only used when computing their fluxes in the detector's energy range while checking for their detectability. Having established that all simulated bursts with $L_\gamma \gtrsim 3\times10^{52}$~erg s$^{-1}$ are detectable regardless of their duration, we can conclude that the concave shape is not affected by the imposed $k$-correction. Instead, it is an intrinsic feature of the model and  result of the convolution of the simulated $\tg$ and $\Lg$ distributions. By populating this part of the $\Lg-\tg$ phase space with more observations in the future, we will be able to test if the model prediction is supported by the data.

Figure~\ref{fig:densmap-Eiso} shows the density maps of detectable (left panel) and successful (right panel) simulated GRBs in the $E_{\gamma}-\Lg$ plane. Here, $E_{\gamma}=\tg \Lg$, is the (bolometric) isotropic gamma-ray energy. The right panel shows that there is an intrinsic correlation of $E_{\gamma}$ and $\Lg$ in the model, with a spread that becomes smaller towards more energetic and luminous bursts (top right corner of the plot). By applying the BAT flux threshold (see equation \ref{eq:minflux}), we find that the density map of detectable simulated bursts in the $E_{\gamma}-\Lg$ plane captures the main features of the observed 2D distribution of long GRBs in this phase space (see left panel).  Similarly to Figure~\ref{fig:densmap}, we find  that the center of the density map of successful bursts moves to lower $E_{\gamma}$ and $\Lg$ values compared with the peak position in the map of detectable bursts.

Figure~\ref{fig:heatmap} shows contour plots of various central engine properties for all detectable bursts from our Monte Carlo simulation (from left to right): $\log (\te/\tg)$,  $\log \tb$, and $\log(E_{\gamma}/E_{\rm e})$. Here, $E_{\rm e}\equiv L_{\rm e} \te$, is the (isotropic equivalent) energy of the central engine. 
Inspection of the plot on the left-hand side shows that $\te \approx \tg \gtrsim 300$~s (with a weak dependence on $\Lg$), suggesting that the  gamma-ray duration distribution at the longest timescales follows closely the distribution of engine timescales, in agreement with previous findings \citep{bromberg2012, PBDG2017}. This follows from the fact that the breakout time (even for the least powerful engines) is shorter than the engine activity timescale for $\tg \gtrsim 300$~s (see middle panel). 
Our model predicts successful engines with much longer activity timescales than the gamma-ray burst duration (as indicated by the color bar on top of the left hand-side panel). The ratio $\te/\tg$ increases as $\tg$ becomes shorter, while it depends only weakly on $\Lg$. Our results suggest that the duration of the gamma-ray emission is not always an indicator of the GRB central engine activity timescale. The colour gradient in the contour plot of $\log \tb$ (central panel) reflects the underlying relation between the engine power and the breakout time (i.e., $\tb\propto L_{\rm e}^{-1/3}$). Therefore, roughly speaking, horizontal cuts in the $\Lg-\tg$ plane pick up GRB jets with similar breakout times. The energy ratio plotted in the rightmost panel of the figure can be interpreted as an efficiency of transforming the central engine's energy into radiated gamma-ray energy, and can be written as $E_{\gamma}/E_{\rm e}\equiv \eta_{\gamma} (t_{\gamma}/t_{\rm e})$. Given that $\eta_{\gamma}$ for detectable bursts has an almost uniform distribution between 0.01 and 0.1 (see also Figure~\ref{fig:histo}), the ratio $E_{\gamma}/E_{\rm e}$ can be mapped to the ratio of the respective timescales. This also explains why the colour map of the energy ratio is (almost) the inverse of the colour map of the timescales ratio. Thus, diagonal cuts in the $\Lg-\tg$ plane probe bursts with similar efficiencies in converting engine energy into gamma-rays. 

\section{Discussion}\label{sec:discuss}
In this section, we first examine potential caveats in our analysis and then move to discuss the implications of our simulation results.
\subsection{Caveats}\label{sec:caveats}
\textit{Completeness of the collapsar sample.} When building our observed GRB sample, we implicitly assumed that  bursts with $T_{90}\ge2$~s are collapsars. In reality, there is not a one-to-one correspondence between long (short) GRBs and collapsars (non-collapsars). \citet{bromberg2013} estimated that $\sim40\%$ of \swift \, bursts with $T_{90}<2$~s are likely to be collapsars despite their short duration. Based on the above, it is likely that $\sim8$ bursts from our short GRB sample are also collapsars, and as such, should be included in the parameter exploration (see Section~\ref{sec:MC-param}). Still, the addition of $\sim\mathcal{O}(10)$ more bursts to our initial sample of 291 collapsar GRBs should not affect the results of the parameter exploration. Using the results of the Monte Carlo simulation presented in Section~\ref{sec:results}, we created $10^4$ random sub-samples of detectable bursts, each having size equal to the size of our combined sample of observed short and long GRBs,  and computed the fraction of (detectable) collapsars  with $T_{90}<2$~s that we would missclassify as non-collapsars because of the adopted threshold in duration. We find that the 68\% of the fraction values lies between $\sim5\%$ and 30\%, which is consistent (within uncertainties) with the findings of \cite{bromberg2013}. 

\textit{Bolometric gamma-ray luminosity.}  
When comparing the samples of simulated and observed bursts, we used the bolometric gamma-ray luminosity. As explained in Section~\ref{sec:data}, we use the best-fit spectral parameters from the third BAT GRB catalog, which were derived by fitting the observed spectra in the detector's energy range. We acknowledge that by extrapolating the spectral fits, in particularly hard power-law spectra, beyond the BAT energy range we may overestimate the bolometric luminosity for certain bursts (e.g., because the turn-over point in the spectrum might happen somewhere above the BAT energy limit). 
Given that we do not search for the best-fit model parameters
(for details, see Section~\ref{sec:MC-param}), the uncertainty in estimating the bolometric gamma-ray luminosity of some bursts will not alter the main conclusions of this study.  

Nevertheless, we tested the soundness of our bolometric estimates as follows. First, we compared how the bolometric isotropic energies we derived in Section~\ref{sec:data} compare with those published in other studies for matching bursts. Using the best-fit spectral parameters of the Band function \citep{Band1993} listed in Table~2 of \cite{Li2016}, we computed the isotropic bolometric energy for 37 \swift-BAT long GRBs that we have in common. Although on a burst-to-burst basis there are differences (factor of $\sim1.5-2$), we find no systematic differences when considering the whole sample. Given that the choice of the spectral model does not introduce systematic uncertainties in the estimation of bolometric quantities \citep[for a discussion on spectral models see][and references therein]{3rdBAT}, this choice is not expected to affect the main results of our simulations (see Sections~\ref{sec:MC} and \ref{sec:results}). We also considered the isotropic bolometric burst energies presented in \cite{Butler2010}. These authors estimated the isotropic energies of \swift-BAT bursts using a Bayesian approach for the spectral fitting \citep[for details, see][]{Butler2007}, which incorporates into the priors for the model parameters knowledge from a large number of pre-\swift \, observations \citep{Preece2000}. We compared the $E_{\gamma}$ estimates for our long GRB sample with those for 67 bursts with confirmed redshift from \cite{Butler2010},  and found that our method yields on average $\sim 3$ higher isotropic energies than those reported in \cite{Butler2010}. Given this systematic deviation (which may stem from differences in the spectral fitting method), we also investigated how the results of our parameter exploration would change if our method led to a systematic overestimation of the true $\Lg$ for all bursts in the sample by a factor of 3. The comparison of the model to the data, after correcting for the overestimation of $\Lg$, yields qualitatively similar results as those presented in Section~\ref{sec:MC}, while the distributions of $\eta_{\gamma,\min}$ and $L_{\rm e, min}$ for ``acceptable'' cases are shifted by a similar factor to lower and higher values, respectively. A smaller shift in $\alpha$ ($\sim15\%$)  is also found. 

We also applied the $k$-correction to transform the bolometric fluxes of simulated GRBs to fluxes in the detector's energy range while checking their detectability (see Section~\ref{sec:MC-grb}). The fact that we are using the same method for estimating the bolometric luminosities of \swift-BAT GRBs and simulated GRBs, suggests that, if there are any biases in our estimation, these will equally affect both samples. In conclusion, systematic differences (of a factor of a few) in the estimation of bolometric quantities  (of simulated and observed bursts) can affect the specific range of ``acceptable'' parameter values, but they will not alter the main conclusion of our study: central engines with unrelated power and activity timescales can account for the main features of the observed 2D distribution of long GRBs in the $\Lg-\tg$ plane, when taking into the account the dependence of the jet breakout timescale on the engine power and the effects of the detector's flux threshold.

\textit{GRB simulations.} The computation of $\Lg$ for the simulated bursts is crude, as it does not take into account the varying spectral properties (e.g., evolution of peak energy and peak luminosity) and pulse profiles of observed GRBs. Moreover, our analysis is not designed to  take into account the energy-dependence of $\tg$, as found in a study of gamma-ray bursts detected by the \emph{Fermi} Gamma-ray Burst Monitor \citep{2013ApJ...763...15Q}. In principle, time-resolved photon spectra for each simulated burst with a time-averaged $\Lg$ and characteristic duration $\tg$ (for a specific energy band) should be computed. To assess whether a simulated burst would have been detected by BAT in this case, count-rate light curves should be computed by folding the time-resolved photon spectra to the instrument's response matrix \citep[see e.g.,][]{kocevski2012}. However, such simulations are disproportionate to the scope of this work, given that we are not trying to fit the model to the data, as explained in Section~\ref{sec:MC-param}. 

\textit{Model-data comparison.} In Section~\ref{sec:MC-param} we used a simplified Monte Carlo scheme to investigate the six-dimensional parameter space, having some prior knowledge for the parameter values from the analytical study of \cite{PBDG2017}. We refrained from using a Markov Chain Monte Carlo (MCMC) method for the sampling of multi-dimensional probability distributions \citep[e.g.,][]{2004PhRvD..69j3501T}, as there is some randomness encoded in the outcomes of the underlying model\footnote{We generate our samples through Monte Carlo simulations, and we find some variance in the properties of the generated bursts even for the same parameter values.}, which complicates the definition of a likelihood function. Moreover, the construction of a large sample of successful simulated bursts for certain parameter combinations can be challenging and may stall the Monte Carlo simulation itself (see Section~\ref{sec:MC-grb}). 
An MCMC approach would be best suited for performing a wide search of the multi-parameter space, if a 2D probability density function (i.e., $p(\Lg,\tg) {\rm d}\Lg {\rm d}\tg$) for detectable bursts could be analytically constructed, but this warrants a separate study.

\begin{figure}
    \centering
    \includegraphics[width=0.45\textwidth]{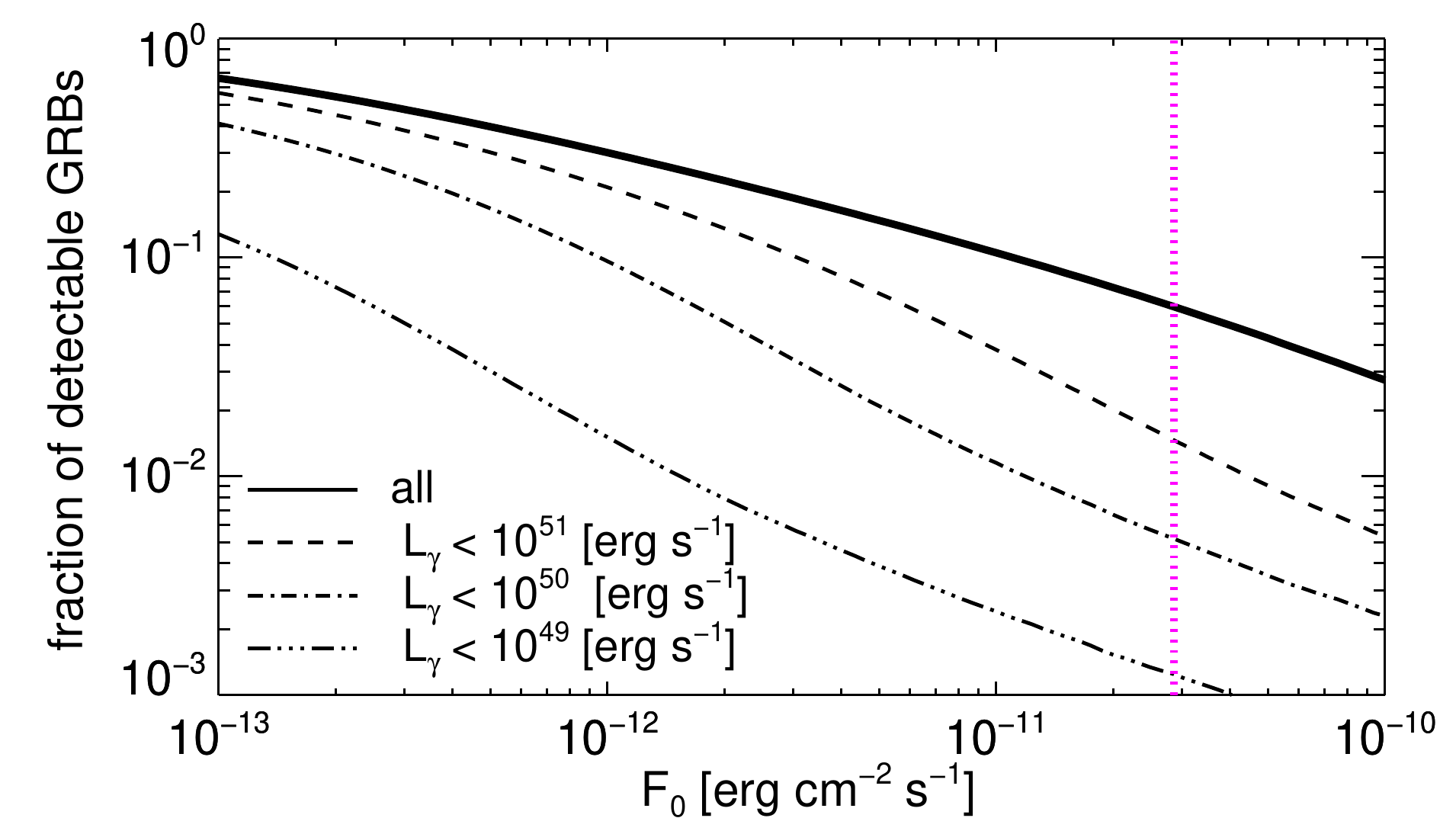}
    \caption{Fraction of simulated detectable GRBs as a function of the flux threshold $F_0$ of a hypothetical detector with limiting flux given by equation~(\ref{eq:minflux}). The fraction of detectable GRBs below a certain value of $\Lg$ is also shown (for details, see inset legend). For comparison, we also show the flux threshold value of \swift-BAT (vertical magenta line).}
    \label{fig:flim}
\end{figure}

\subsection{Flux threshold}\label{sec:sensitivity}
One of our main model predictions is that the intrinsic distribution of collapsar GRBs in the $\Lg-\tg$ plane should peak to lower isotropic luminosities and longer durations compared with the observed sample of \swift-BAT bursts (see Figure~\ref{fig:densmap}). It is therefore interesting to check how the number of detectable bursts  from our Monte Carlo simulation depends on the flux threshold of a detector which, in all other aspects besides sensitivity, is assumed to similar to \swift-BAT  (e.g., energy range and triggering method). Our results are presented in Figure~\ref{fig:flim}. A three (thirty) times more sensitive detector than \swift-BAT would yield $\sim1.8$ ($\sim 5$) more detectable bursts, while the fraction of detectable bursts with $\Lg<10^{50}$~erg~s$^{-1}$ would increase by a factor of $\sim2$ ($\sim 22$). Although the exact numbers are subject to the detector's specific capabilities, the model's prediction about a faster increase in the fraction of detectable bursts with lower luminosities (and longer rest-frame durations) is robust. 

\subsection{Universal radiative efficiency?}
\label{sec:eta}
We have presented results obtained under the assumption that the intrinsic distribution of $\log \etag $ is extended and, in particular, uniform (see Figure~\ref{fig:histo}). Although we find reasonable agreement between the observed and simulated (detectable) samples for the chosen range of $\etag$ values, we remind that $\eta_{\gamma,\min}$ could not be constrained in the parameter exploration (see Figure~\ref{fig:mcmc}), while the upper bound of the distribution was fixed. It is therefore reasonable to test if the choice of a common $\etag$ for all simulated bursts would also yield a sample of $\Lg,\tg$ values compatible with the observed one. To do so, we performed another Monte Carlo simulation of $10^6$ successful GRBs, assuming $\etag=0.01$ or $\etag=0.1$, while keeping all other parameters the same as before (see Table~\ref{tab:param}). For each case, we constructed the density map of detectable bursts in the $\Lg-\tg$ plane and performed a 2D KS test between a random sub-sample of 1000 detectable bursts and the observed sample of GRBs. We found the following ordering for the test statistic values, $Z_{\rm n,2D}^{(\etag=0.01)} \gg Z_{\rm n,2D}^{(\etag=0.1)} \approx Z_{\rm n,2D}^{\rm (uni)}$. For $\etag=0.01$, one can tell even by eye that the two samples in the $\Lg-\tg$ plane are different, as the model fails to reproduce bursts with $\Lg \gtrsim 10^{52}$~erg s$^{-1}$ (not shown here). Although, currently we cannot distinguish between scenarios with $\etag \sim 0.1$  and a uniform distribution of efficiencies ($\sim 0.01-0.25$), we should be able to do so in the future with more sensitive detectors. 
The former scenario predicts a very sharp cutoff at $\etag L_{\rm e,min}$ independent of the GRB duration, while the latter predicts a smoother cutoff with some dependence on $\tg$, as shown in the right panel of Figure~\ref{fig:densmap}.

\subsection{Universal breakout-time parameter?}\label{sec:t0}
We have performed our simulations under the assumption of a common breakout-time parameter, $t_0$, for all bursts. Still, some variance within the GRB population is expected, as $t_0$ encodes information about the progenitor's properties and the jet injection angle \citep[see e.g., equation 1 in][]{bromberg2012}. Given the weak dependence of $t_0$ on the stellar properties ($t_0 \propto M_{*}^{1/3} R_{*}^{2/3}$) and the fact that these are not expected to vary by orders of magnitude among collapsars, their effect on $t_0$ should be negligible. The only parameter that could then introduce scatter in the $t_0$ values of the GRB population is the jet injection angle $\theta_0$. \cite{Goldstein2016}, for example, estimated the jet opening angles of long GRBs, and found that these follow a log-normal distribution (with 90\% of the values lying below 20 degrees). Assuming that the injection angles have a similar distribution, then $t_0$  should also follow a log-normal distribution with smaller spread around the mean ($t_0 \propto \theta_0^{2/3}$). 

Motivated by this, we relaxed our prior assumption on $t_0$ and considered a case where $\log(t_0)$ of each simulated burst is sampled from a normal distribution with mean $\mu_{t_0}$ and standard deviation $\sigma_{t_0}$. We then repeated the parameter exploration, as described in Section~\ref{sec:MC-param}, with an additional free parameter and $\mu_{t_0}$ replacing $t_0$. To simplify the parameter scan (and because $L_{\rm e,0}$ is correlated with $t_0$), we fixed $L_{\rm e,0}$ to the value listed in Table~\ref{tab:param}. The median values of  $\mu_{t_0}$ and $\sigma_{t_0}$ for the acceptable parameter combinations read $2.09$ and $0.017$, respectively. Meanwhile the median values (and the 68\% intervals) of all other parameters are very similar to those listed in Table~\ref{tab:param}. The results of this example suggest that a narrow distribution of $t_0$ values clustered around $\sim 120$~s is required by our empirical central engine model. 

As pointed out by \cite{PBDG2017}, values  of $t_0 \gg 20$~s cannot be easily reconciled with the scenario of jet propagation through a compact progenitor, because of the weak dependence that $t_0$ has on stellar properties \citep{bromberg2011, bromberg2012}. Interestingly, we reach to the same conclusion after performing a 2D comparison of the model to the data. Thus, our empirical model for the engines of GRB collapsars implies the presence of an extended low-mass envelope surrounding the GRB progenitor, as independently concluded by \cite{sobacchi2017}.
 
\subsection{Engine activity timescale}
Our model for the central engine is based upon the simple equation $\tg=\te-\tb(\Le)$ that relates the gamma-ray duration of the prompt emission $\tg$ to the jet breakout time $\tb$ and the engine activity timescale $\te$. By construction, the latter refers to the engine activity time in which the bulk of the burst radiation is released, i.e., the prompt GRB signal. During this time the engine power, $\Le$, is also assumed to be constant. The detection of late-time X-ray flares (and X-ray plateaus) may indicate that the central engine remains active or restarts, albeit at a reduced luminosity and total energy, long after the main GRB episode in over (e.g., \citealt{2005Sci...309.1833B, King2005,Liang2006,Troja2007, 2013ApJ...763...15Q, 2014ApJ...787...66Z}; see, however, \citealt{BK2016,BDDM2020}). Our analysis, however, is not meant to describe this late-time tail of the engine activity. Thus, the inferred values for $\te$ should be considered as lower limits of the true engine duration in collapsars exhibiting late-time X-ray activity.  

\subsection{Nature of the central engine}
Our generic central engine model is built upon the assumption that the engine luminosity and activity timescale are independent of each other. This hypothesis yields results that are so far consistent with the 2D distribution of observed long GRBs in the $\Lg-\tg$ and $E_{\gamma}-\Lg$ planes (see Figures~\ref{fig:densmap} and \ref{fig:densmap-Eiso}). If this hypothesis is also confirmed by more detailed simulations and/or larger data sets, it will have important implications for the central engine of collapsars. 

If the central engine is a black hole, accretion in combination to the rotation energy of the black hole can power the GRBs \citep[e.g.,][]{Meszaros1997}. The in-falling stellar material can drag in the large-scale magnetic flux through the progenitor star and the jet can be powered via the Blandford-Znajek  process \citep{BZ1977}. In this scenario, the jet power (equivalent to the power of the central engine) is determined by the magnetic flux through the BH horizon $\Phi_{\rm BH}$, namely $L_{\rm e} \equiv L_{\rm BZ} \propto a_{\rm BH} \Phi^2_{\rm BH} M^{-2}_{\rm BH}$, where $a_{\rm BH}$ and  $M_{\rm BH}$ are the BH spin and mass, respectively. As long as the accretion rate $\dot{M}$ is high enough as to sustain the magnetic flux $\Phi_{\rm BH}$ on the BH, the jet power is independent of $\dot{M}$ and approximately constant. However, as $\dot{M}$ decreases with time after core collapse, at some point the gas pressure in the disk becomes too weak to hold the magnetic flux on the BH~(\cite{TG15} henceforth, TG15). This happens at a critical mass accretion rate where $\dot{M}_{\rm MAD}c^2 \approx L_{\rm BZ}$. Soon after the accretion rate drops below $\dot{M}_{\rm MAD}$, part of the magnetic flux diffuses out, while the remaining flux to the BH inhibits gas accretion, leading to the formation of a magnetically arrested disk \citep[MAD,][]{narayan2003, 2011MNRAS.418L..79T, 2012MNRAS.423.3083M}. In this scenario, the times of the accretion disk formation and the MAD onset   define the engine activity timescale $\te$ \citep{TG15}. TG15 studied the dependence of $\Le$ and $\te$ on several parameters, including the magnetic flux and stellar progenitor model. To a first approximation, the engine power is mainly set by $\Phi_{\rm BH}$, which has only a very weak effect on $\te$; a change of 1000 in magnetic flux results  almost in the same engine time (see Figure 6 in TG15). The duration $\te$ is mainly affected by other properties of the progenitor, such as stellar rotation rate (see Figure 10 in TG15) or the progenitor model itself (see Figure 15 in TG15). Unless the magnetic flux is tied to the progenitor properties (e.g., faster rotating progenitors have larger magnetic flux), this model can explain engines with $\Le$ and $\te$ independent of each other.

In the ``vanilla" magnetar model for the central engine, where the jet is solely powered  by the solid-body spin-down energy of the magnetar, the engine luminosity can be written as  $L_{\rm e}\propto L_{\rm SD}\propto \Omega^4 B^2$, where $\Omega$ is  the initial spin frequency and $B$ the surface magnetic field. The engine duration is $t_{\rm e}=\min(t_{\rm SD},t_{\sigma_0})$ \citep{BGM2017}, where $t_{\rm SD}\propto \Omega^{-2} B^{-2}$ is the magnetar spin-down timescale, and $t_{\sigma_0}\approx 100\mbox{ s}$ is the time it takes for the jet to turn to an essentially baryon-free pulsar wind (this timescale is not expected to vary much between bursts). Assuming that $\Omega,B$ are independent of each other, this model predicts an anti-correlation between $L_{\rm e}$ and $t_{\rm e}$ for engine durations $\lesssim 100\mbox{ s}$, and a very sharp cutoff of the duration distribution at $\tg \sim 100$~s. The predicted scaling relation ranges between $L_{\rm e}\propto t_{\rm e}^{-2}$ for varying $\Omega$ and constant $B$, and $L_{\rm e}\propto t_{\rm e}^{-1}$ for the extreme opposite case. Anti-correlations in the properties of the central engine generally lead to strong anti-correlations between the observable quantities. Unless there is enough scatter in the intrinsic model parameters, the predicted $\Lg-\tg$ distribution is narrow and in tension with the observed data. A dedicated study of the magnetar model, including also the effects of the fallback accretion \citep{MBG2018}, will be the topic of a future study.

One of the motivations for the magnetar central engine model is the detection of X-ray plateaus and late-time flares in the GRB afterglow light curves, which have been attributed to late-time energy injection from a magnetar \citep[see][and references therein]{Stratta2018}. At the same time, GRBs with very large gamma-ray isotropic energies and/or prompt emission durations, are not readily accountable by the magnetar model \citep[e.g.,][]{Cenko2010, Li2018}. In contrast to black-hole engines, the available energy to power a GRB is limited by the magnetar's rotational energy ($\sim 2\times10^{52}$~erg) \citep[e.g.,][]{Usov1992, Thompson2004}. The requirement of producing a relativistic outflow and the uncertain radiative efficiency can further limit the available energy to power a GRB in magnetar models \citep[e.g.,][]{Metzger+11, BGM2017}.
One may postulate then, that two populations of long GRBs exist, i.e., those powered by magnetars and those powered by black holes \citep[e.g.,][]{Li2018}. Even if this is the case, it is not clear if these populations may be discernible from prompt emission observations. Here, we test the hypothesis that the $\Lg-\tg$ samples of BAT bursts with and without plateaus in their \swift-XRT light curves belong to the same parent population. We use the sample of bursts with plateaus given by \citet{Tang2019} and split our sample of 291 long GRBs (see Section~\ref{sec:data}) to two samples composed of bursts with and without X-ray plateau. We find no statistically significant difference between the samples, and the null hypothesis, that they are drawn from the same underlying population, cannot be ruled out. 
Thus, even if bursts with and without X-ray plateaus in their afterglows are powered by different types of central engines, their prompt emission properties (i.e., $\Lg$ and $\tg$) cannot be used to distinguish between the two populations.

\section{Conclusions}\label{sec:conclusions}
Jets in long GRBs have to drill through the collapsing star in order to break out of it and produce the gamma-ray signal while the central engine is still active. Using Monte Carlo simulations that take into account the dependence of the jet breakout timescale on the engine luminosity and the effects of the detector's flux threshold, we showed that central engines with unrelated luminosities and activity timescales 
can reproduce the main features of the 2D distribution of \swift-BAT long GRBs in the $\Lg-\tg$ plane. According to our model, the intrinsic 2D distribution of collapsar GRBs peaks at lower gamma-ray luminosities and longer durations than the observed one, a prediction that can be tested in the future with more sensitive detectors.

\section*{Acknowledgements}
The authors thank the anonymous referee for constructive comments. The authors thank Dr. J. Buchner for useful discussions and comments on the manuscript. M.P. acknowledges support from the Lyman Jr.~Spitzer Postdoctoral Fellowship and the Fermi Guest Investigation grant 80NSSC18K1745. P.B. acknowledges support from the Gordon and Betty Moore Foundation through Grant GBMF5076. R.B.D and D.G. acknowledge support from the National Science Foundation under Grants 1816694 and 1816136. D.G. acknowledges support from the NASA grant NNX17AG21G and the Fermi 
Guest  Investigator Program Cycle 12, grant 80NSSC19K1506.




\bibliographystyle{mnras}
\bibliography{grb} 

\begin{thebibliography}{}
\makeatletter
\relax
\def\mn@urlcharsother{\let\do\@makeother \do\$\do\&\do\#\do\^\do\_\do\%\do\~}
\def\mn@doi{\begingroup\mn@urlcharsother \@ifnextchar [ {\mn@doi@}
  {\mn@doi@[]}}
\def\mn@doi@[#1]#2{\def\@tempa{#1}\ifx\@tempa\@empty \href
  {http://dx.doi.org/#2} {doi:#2}\else \href {http://dx.doi.org/#2} {#1}\fi
  \endgroup}
\def\mn@eprint#1#2{\mn@eprint@#1:#2::\@nil}
\def\mn@eprint@arXiv#1{\href {http://arxiv.org/abs/#1} {{\tt arXiv:#1}}}
\def\mn@eprint@dblp#1{\href {http://dblp.uni-trier.de/rec/bibtex/#1.xml}
  {dblp:#1}}
\def\mn@eprint@#1:#2:#3:#4\@nil{\def\@tempa {#1}\def\@tempb {#2}\def\@tempc
  {#3}\ifx \@tempc \@empty \let \@tempc \@tempb \let \@tempb \@tempa \fi \ifx
  \@tempb \@empty \def\@tempb {arXiv}\fi \@ifundefined
  {mn@eprint@\@tempb}{\@tempb:\@tempc}{\expandafter \expandafter \csname
  mn@eprint@\@tempb\endcsname \expandafter{\@tempc}}}

\bibitem[\protect\citeauthoryear{{Band} et~al.,}{{Band}
  et~al.}{1993}]{Band1993}
{Band} D.,  et~al., 1993, \mn@doi [\apj] {10.1086/172995}, \href
  {http://adsabs.harvard.edu/abs/1993ApJ...413..281B} {413, 281}

\bibitem[\protect\citeauthoryear{{Beniamini} \& {Kumar}}{{Beniamini} \&
  {Kumar}}{2016}]{BK2016}
{Beniamini} P.,  {Kumar} P.,  2016, \mn@doi [\mnras] {10.1093/mnrasl/slw003},
  \href {http://cdsads.u-strasbg.fr/abs/2016MNRAS.457L.108B} {457, L108}

\bibitem[\protect\citeauthoryear{{Beniamini} \& {Mochkovitch}}{{Beniamini} \&
  {Mochkovitch}}{2017}]{BM2017}
{Beniamini} P.,  {Mochkovitch} R.,  2017, \mn@doi [\aap]
  {10.1051/0004-6361/201730523}, \href
  {http://cdsads.u-strasbg.fr/abs/2017A%26A...605A..60B} {605, A60}

\bibitem[\protect\citeauthoryear{{Beniamini}, {Nava}, {Duran}  \&
  {Piran}}{{Beniamini} et~al.}{2015}]{Beniamini2015}
{Beniamini} P.,  {Nava} L.,  {Duran} R.~B.,   {Piran} T.,  2015, \mn@doi
  [\mnras] {10.1093/mnras/stv2033}, \href
  {https://ui.adsabs.harvard.edu/abs/2015MNRAS.454.1073B} {454, 1073}

\bibitem[\protect\citeauthoryear{{Beniamini}, {Nava}  \& {Piran}}{{Beniamini}
  et~al.}{2016}]{Beniamini2016}
{Beniamini} P.,  {Nava} L.,   {Piran} T.,  2016, \mn@doi [\mnras]
  {10.1093/mnras/stw1331}, \href
  {https://ui.adsabs.harvard.edu/abs/2016MNRAS.461...51B} {461, 51}

\bibitem[\protect\citeauthoryear{{Beniamini}, {Giannios}  \&
  {Metzger}}{{Beniamini} et~al.}{2017}]{BGM2017}
{Beniamini} P.,  {Giannios} D.,   {Metzger} B.~D.,  2017, \mn@doi [\mnras]
  {10.1093/mnras/stx2095}, \href
  {http://cdsads.u-strasbg.fr/abs/2017MNRAS.472.3058B} {472, 3058}

\bibitem[\protect\citeauthoryear{{Beniamini}, {Barniol Duran}, {Petropoulou}
  \& {Giannios}}{{Beniamini} et~al.}{2020a}]{BBPG2020}
{Beniamini} P.,  {Barniol Duran} R.,  {Petropoulou} M.,   {Giannios} D.,
  2020a, arXiv e-prints, \href
  {https://ui.adsabs.harvard.edu/abs/2020arXiv200100950B} {p. arXiv:2001.00950}

\bibitem[\protect\citeauthoryear{{Beniamini}, {Duque}, {Daigne}  \&
  {Mochkovitch}}{{Beniamini} et~al.}{2020b}]{BDDM2020}
{Beniamini} P.,  {Duque} R.,  {Daigne} F.,   {Mochkovitch} R.,  2020b, \mn@doi
  [\mnras] {10.1093/mnras/staa070}, \href
  {https://ui.adsabs.harvard.edu/abs/2020MNRAS.492.2847B} {492, 2847}

\bibitem[\protect\citeauthoryear{{Bennett}, {Larson}, {Weiland}  \&
  {Hinshaw}}{{Bennett} et~al.}{2014}]{Bennett2014}
{Bennett} C.~L.,  {Larson} D.,  {Weiland} J.~L.,   {Hinshaw} G.,  2014, \mn@doi
  [\apj] {10.1088/0004-637X/794/2/135}, \href
  {https://ui.adsabs.harvard.edu/abs/2014ApJ...794..135B} {794, 135}

\bibitem[\protect\citeauthoryear{{Blandford} \& {Znajek}}{{Blandford} \&
  {Znajek}}{1977}]{BZ1977}
{Blandford} R.~D.,  {Znajek} R.~L.,  1977, \mn@doi [\mnras]
  {10.1093/mnras/179.3.433}, \href
  {https://ui.adsabs.harvard.edu/abs/1977MNRAS.179..433B} {179, 433}

\bibitem[\protect\citeauthoryear{{Bloom}, {Frail}  \& {Sari}}{{Bloom}
  et~al.}{2001}]{Bloom2001}
{Bloom} J.~S.,  {Frail} D.~A.,   {Sari} R.,  2001, \mn@doi [\aj]
  {10.1086/321093}, \href
  {https://ui.adsabs.harvard.edu/abs/2001AJ....121.2879B} {121, 2879}

\bibitem[\protect\citeauthoryear{{Bromberg}, {Nakar}  \& {Piran}}{{Bromberg}
  et~al.}{2011a}]{bromberg2011}
{Bromberg} O.,  {Nakar} E.,   {Piran} T.,  2011a, \mn@doi [\apjl]
  {10.1088/2041-8205/739/2/L55}, \href
  {http://adsabs.harvard.edu/abs/2011ApJ...739L..55B} {739, L55}

\bibitem[\protect\citeauthoryear{{Bromberg}, {Nakar}, {Piran}  \&
  {Sari}}{{Bromberg} et~al.}{2011b}]{bromberg2011b}
{Bromberg} O.,  {Nakar} E.,  {Piran} T.,   {Sari} R.,  2011b, \mn@doi [\apj]
  {10.1088/0004-637X/740/2/100}, \href
  {http://adsabs.harvard.edu/abs/2011ApJ...740..100B} {740, 100}

\bibitem[\protect\citeauthoryear{{Bromberg}, {Nakar}, {Piran}  \&
  {Sari}}{{Bromberg} et~al.}{2012}]{bromberg2012}
{Bromberg} O.,  {Nakar} E.,  {Piran} T.,   {Sari} R.,  2012, \mn@doi [\apj]
  {10.1088/0004-637X/749/2/110}, \href
  {http://adsabs.harvard.edu/abs/2012ApJ...749..110B} {749, 110}

\bibitem[\protect\citeauthoryear{{Bromberg}, {Nakar}, {Piran}  \&
  {Sari}}{{Bromberg} et~al.}{2013}]{bromberg2013}
{Bromberg} O.,  {Nakar} E.,  {Piran} T.,   {Sari} R.,  2013, \mn@doi [\apj]
  {10.1088/0004-637X/764/2/179}, \href
  {http://adsabs.harvard.edu/abs/2013ApJ...764..179B} {764, 179}

\bibitem[\protect\citeauthoryear{{Burrows} et~al.,}{{Burrows}
  et~al.}{2005}]{2005Sci...309.1833B}
{Burrows} D.~N.,  et~al., 2005, \mn@doi [Science] {10.1126/science.1116168},
  \href {https://ui.adsabs.harvard.edu/abs/2005Sci...309.1833B} {309, 1833}

\bibitem[\protect\citeauthoryear{{Butler}, {Kocevski}, {Bloom}  \&
  {Curtis}}{{Butler} et~al.}{2007}]{Butler2007}
{Butler} N.~R.,  {Kocevski} D.,  {Bloom} J.~S.,   {Curtis} J.~L.,  2007,
  \mn@doi [\apj] {10.1086/522492}, \href
  {https://ui.adsabs.harvard.edu/abs/2007ApJ...671..656B} {671, 656}

\bibitem[\protect\citeauthoryear{{Butler}, {Bloom}  \& {Poznanski}}{{Butler}
  et~al.}{2010}]{Butler2010}
{Butler} N.~R.,  {Bloom} J.~S.,   {Poznanski} D.,  2010, \mn@doi [\apj]
  {10.1088/0004-637X/711/1/495}, \href
  {https://ui.adsabs.harvard.edu/abs/2010ApJ...711..495B} {711, 495}

\bibitem[\protect\citeauthoryear{{Cenko} et~al.,}{{Cenko}
  et~al.}{2010}]{Cenko2010}
{Cenko} S.~B.,  et~al., 2010, \mn@doi [\apj] {10.1088/0004-637X/711/2/641},
  \href {https://ui.adsabs.harvard.edu/abs/2010ApJ...711..641C} {711, 641}

\bibitem[\protect\citeauthoryear{{Chen} \& {Beloborodov}}{{Chen} \&
  {Beloborodov}}{2007}]{chen2007}
{Chen} W.-X.,  {Beloborodov} A.~M.,  2007, \mn@doi [\apj] {10.1086/508923},
  \href {https://ui.adsabs.harvard.edu/abs/2007ApJ...657..383C} {657, 383}

\bibitem[\protect\citeauthoryear{{Dai}, {Wang}, {Wu}  \& {Zhang}}{{Dai}
  et~al.}{2006}]{Dai2006}
{Dai} Z.~G.,  {Wang} X.~Y.,  {Wu} X.~F.,   {Zhang} B.,  2006, \mn@doi [Science]
  {10.1126/science.1123606}, \href
  {http://cdsads.u-strasbg.fr/abs/2006Sci...311.1127D} {311, 1127}

\bibitem[\protect\citeauthoryear{{Eichler}, {Livio}, {Piran}  \&
  {Schramm}}{{Eichler} et~al.}{1989}]{Eichler1989}
{Eichler} D.,  {Livio} M.,  {Piran} T.,   {Schramm} D.~N.,  1989, \mn@doi
  [\nat] {10.1038/340126a0}, \href
  {https://ui.adsabs.harvard.edu/abs/1989Natur.340..126E} {340, 126}

\bibitem[\protect\citeauthoryear{{Fasano} \& {Franceschini}}{{Fasano} \&
  {Franceschini}}{1987}]{FF1987}
{Fasano} G.,  {Franceschini} A.,  1987, \mn@doi [\mnras]
  {10.1093/mnras/225.1.155}, \href
  {https://ui.adsabs.harvard.edu/abs/1987MNRAS.225..155F} {225, 155}

\bibitem[\protect\citeauthoryear{{Gehrels} et~al.,}{{Gehrels}
  et~al.}{2004}]{Gehrels2004}
{Gehrels} N.,  et~al., 2004, \mn@doi [\apj] {10.1086/422091}, \href
  {http://adsabs.harvard.edu/abs/2004ApJ...611.1005G} {611, 1005}

\bibitem[\protect\citeauthoryear{{George}, {Fabian}, {Baumgartner}, {Mushotzky}
   \& {Tueller}}{{George} et~al.}{2008}]{George2008}
{George} M.~R.,  {Fabian} A.~C.,  {Baumgartner} W.~H.,  {Mushotzky} R.~F.,
  {Tueller} J.,  2008, \mn@doi [\mnras] {10.1111/j.1745-3933.2008.00499.x},
  \href {https://ui.adsabs.harvard.edu/abs/2008MNRAS.388L..59G} {388, L59}

\bibitem[\protect\citeauthoryear{{Ghisellini}, {Ghirlanda}, {Tavecchio},
  {Fraternali}  \& {Pareschi}}{{Ghisellini} et~al.}{2008}]{Ghisellini2008}
{Ghisellini} G.,  {Ghirlanda} G.,  {Tavecchio} F.,  {Fraternali} F.,
  {Pareschi} G.,  2008, \mn@doi [\mnras] {10.1111/j.1745-3933.2008.00547.x},
  \href {https://ui.adsabs.harvard.edu/abs/2008MNRAS.390L..88G} {390, L88}

\bibitem[\protect\citeauthoryear{{Goldstein}, {Connaughton}, {Briggs}  \&
  {Burns}}{{Goldstein} et~al.}{2016}]{Goldstein2016}
{Goldstein} A.,  {Connaughton} V.,  {Briggs} M.~S.,   {Burns} E.,  2016,
  \mn@doi [\apj] {10.3847/0004-637X/818/1/18}, \href
  {https://ui.adsabs.harvard.edu/abs/2016ApJ...818...18G} {818, 18}

\bibitem[\protect\citeauthoryear{{Harari}, {Mollerach}  \& {Roulet}}{{Harari}
  et~al.}{2009}]{Harari2009}
{Harari} D.,  {Mollerach} S.,   {Roulet} E.,  2009, \mn@doi [\mnras]
  {10.1111/j.1365-2966.2008.14327.x}, \href
  {https://ui.adsabs.harvard.edu/abs/2009MNRAS.394..916H} {394, 916}

\bibitem[\protect\citeauthoryear{{Hjorth} et~al.}{{Hjorth}
  et~al.}{2003}]{Hjorth+03}
{Hjorth} J.,  et~al., 2003, \mn@doi [\nat] {10.1038/nature01750}, \href
  {http://adsabs.harvard.edu/abs/2003Natur.423..847H} {423, 847}

\bibitem[\protect\citeauthoryear{{Kawanaka}, {Piran}  \& {Krolik}}{{Kawanaka}
  et~al.}{2013}]{Kawanaka2013}
{Kawanaka} N.,  {Piran} T.,   {Krolik} J.~H.,  2013, \mn@doi [\apj]
  {10.1088/0004-637X/766/1/31}, \href
  {https://ui.adsabs.harvard.edu/abs/2013ApJ...766...31K} {766, 31}

\bibitem[\protect\citeauthoryear{{King}, {O'Brien}, {Goad}, {Osborne}, {Olsson}
   \& {Page}}{{King} et~al.}{2005}]{King2005}
{King} A.,  {O'Brien} P.~T.,  {Goad} M.~R.,  {Osborne} J.,  {Olsson} E.,
  {Page} K.,  2005, \mn@doi [\apjl] {10.1086/496881}, \href
  {http://cdsads.u-strasbg.fr/abs/2005ApJ...630L.113K} {630, L113}

\bibitem[\protect\citeauthoryear{{Klu{\'z}niak} \& {Ruderman}}{{Klu{\'z}niak}
  \& {Ruderman}}{1998}]{Kluzniak1998}
{Klu{\'z}niak} W.,  {Ruderman} M.,  1998, \mn@doi [\apjl] {10.1086/311622},
  \href {http://cdsads.u-strasbg.fr/abs/1998ApJ...505L.113K} {505, L113}

\bibitem[\protect\citeauthoryear{{Kocevski}}{{Kocevski}}{2012}]{kocevski2012}
{Kocevski} D.,  2012, \mn@doi [\apj] {10.1088/0004-637X/747/2/146}, \href
  {https://ui.adsabs.harvard.edu/abs/2012ApJ...747..146K} {747, 146}

\bibitem[\protect\citeauthoryear{{Kumar} \& {Zhang}}{{Kumar} \&
  {Zhang}}{2015}]{kumarandzhang2015}
{Kumar} P.,  {Zhang} B.,  2015, \mn@doi [\physrep]
  {10.1016/j.physrep.2014.09.008}, \href
  {http://adsabs.harvard.edu/abs/2015PhR...561....1K} {561, 1}

\bibitem[\protect\citeauthoryear{{Lazzati}, {Morsony}, {Blackwell}  \&
  {Begelman}}{{Lazzati} et~al.}{2012}]{lazzatietal2012}
{Lazzati} D.,  {Morsony} B.~J.,  {Blackwell} C.~H.,   {Begelman} M.~C.,  2012,
  \mn@doi [\apj] {10.1088/0004-637X/750/1/68}, \href
  {http://adsabs.harvard.edu/abs/2012ApJ...750...68L} {750, 68}

\bibitem[\protect\citeauthoryear{{Leng} \& {Giannios}}{{Leng} \&
  {Giannios}}{2014}]{Leng2014}
{Leng} M.,  {Giannios} D.,  2014, \mn@doi [\mnras] {10.1093/mnrasl/slu122},
  \href {https://ui.adsabs.harvard.edu/abs/2014MNRAS.445L...1L} {445, L1}

\bibitem[\protect\citeauthoryear{{Li}, {Zhang}  \& {L{\"u}}}{{Li}
  et~al.}{2016}]{Li2016}
{Li} Y.,  {Zhang} B.,   {L{\"u}} H.-J.,  2016, \mn@doi [\apjs]
  {10.3847/0067-0049/227/1/7}, \href
  {https://ui.adsabs.harvard.edu/abs/2016ApJS..227....7L} {227, 7}

\bibitem[\protect\citeauthoryear{{Li}, {Wu}, {Lei}, {Dai}, {Liang}  \&
  {Ryde}}{{Li} et~al.}{2018}]{Li2018}
{Li} L.,  {Wu} X.-F.,  {Lei} W.-H.,  {Dai} Z.-G.,  {Liang} E.-W.,   {Ryde} F.,
  2018, \mn@doi [\apjs] {10.3847/1538-4365/aabaf3}, \href
  {https://ui.adsabs.harvard.edu/abs/2018ApJS..236...26L} {236, 26}

\bibitem[\protect\citeauthoryear{{Liang} et~al.,}{{Liang}
  et~al.}{2006}]{Liang2006}
{Liang} E.~W.,  et~al., 2006, \mn@doi [\apj] {10.1086/504684}, \href
  {http://adsabs.harvard.edu/abs/2006ApJ...646..351L} {646, 351}

\bibitem[\protect\citeauthoryear{{Lien} et~al.,}{{Lien} et~al.}{2016}]{3rdBAT}
{Lien} A.,  et~al., 2016, \mn@doi [\apj] {10.3847/0004-637X/829/1/7}, \href
  {https://ui.adsabs.harvard.edu/\#abs/2016ApJ...829....7L} {829, 7}

\bibitem[\protect\citeauthoryear{{L{\"u}} \& {Zhang}}{{L{\"u}} \&
  {Zhang}}{2014}]{2014ApJ...785...74L}
{L{\"u}} H.-J.,  {Zhang} B.,  2014, \mn@doi [\apj]
  {10.1088/0004-637X/785/1/74}, \href
  {https://ui.adsabs.harvard.edu/abs/2014ApJ...785...74L} {785, 74}

\bibitem[\protect\citeauthoryear{{MacFadyen} \& {Woosley}}{{MacFadyen} \&
  {Woosley}}{1999}]{macfadyen1999}
{MacFadyen} A.~I.,  {Woosley} S.~E.,  1999, \mn@doi [\apj] {10.1086/307790},
  \href {http://adsabs.harvard.edu/abs/1999ApJ...524..262M} {524, 262}

\bibitem[\protect\citeauthoryear{{MacFadyen}, {Woosley}  \&
  {Heger}}{{MacFadyen} et~al.}{2001}]{macfadyen2001}
{MacFadyen} A.~I.,  {Woosley} S.~E.,   {Heger} A.,  2001, \mn@doi [\apj]
  {10.1086/319698}, \href
  {https://ui.adsabs.harvard.edu/abs/2001ApJ...550..410M} {550, 410}

\bibitem[\protect\citeauthoryear{{McKinney}, {Tchekhovskoy}  \& {Bland
  ford}}{{McKinney} et~al.}{2012}]{2012MNRAS.423.3083M}
{McKinney} J.~C.,  {Tchekhovskoy} A.,   {Bland ford} R.~D.,  2012, \mn@doi
  [\mnras] {10.1111/j.1365-2966.2012.21074.x}, \href
  {https://ui.adsabs.harvard.edu/abs/2012MNRAS.423.3083M} {423, 3083}

\bibitem[\protect\citeauthoryear{{M{\'e}sz{\'a}ros} \&
  {Rees}}{{M{\'e}sz{\'a}ros} \& {Rees}}{1997}]{Meszaros1997}
{M{\'e}sz{\'a}ros} P.,  {Rees} M.~J.,  1997, \mn@doi [\apjl] {10.1086/310692},
  \href {https://ui.adsabs.harvard.edu/abs/1997ApJ...482L..29M} {482, L29}

\bibitem[\protect\citeauthoryear{Metchev \& Grindlay}{Metchev \&
  Grindlay}{2002}]{Metchev2002}
Metchev S.~A.,  Grindlay J.~E.,  2002, \mn@doi [Monthly Notices of the Royal
  Astronomical Society] {10.1046/j.1365-8711.2002.05595.x}, 335, 73

\bibitem[\protect\citeauthoryear{{Metzger}, {Giannios}, {Thompson},
  {Bucciantini}  \& {Quataert}}{{Metzger} et~al.}{2011}]{Metzger+11}
{Metzger} B.~D.,  {Giannios} D.,  {Thompson} T.~A.,  {Bucciantini} N.,
  {Quataert} E.,  2011, \mn@doi [\mnras] {10.1111/j.1365-2966.2011.18280.x},
  \href {http://adsabs.harvard.edu/abs/2011MNRAS.413.2031M} {413, 2031}

\bibitem[\protect\citeauthoryear{{Metzger}, {Beniamini}  \&
  {Giannios}}{{Metzger} et~al.}{2018}]{MBG2018}
{Metzger} B.~D.,  {Beniamini} P.,   {Giannios} D.,  2018, \mn@doi [\apj]
  {10.3847/1538-4357/aab70c}, \href
  {http://cdsads.u-strasbg.fr/abs/2018ApJ...857...95M} {857, 95}

\bibitem[\protect\citeauthoryear{Mizuta \& Aloy}{Mizuta \&
  Aloy}{2009}]{mizuta2009}
Mizuta A.,  Aloy M.~A.,  2009, The Astrophysical Journal, 699, 1261

\bibitem[\protect\citeauthoryear{Morsony, Lazzati  \& Begelman}{Morsony
  et~al.}{2007}]{morsony2007}
Morsony B.~J.,  Lazzati D.,   Begelman M.~C.,  2007, The Astrophysical Journal,
  665, 569

\bibitem[\protect\citeauthoryear{{Nakar}}{{Nakar}}{2015}]{nakar2015}
{Nakar} E.,  2015, \mn@doi [\apj] {10.1088/0004-637X/807/2/172}, \href
  {http://adsabs.harvard.edu/abs/2015ApJ...807..172N} {807, 172}

\bibitem[\protect\citeauthoryear{{Narayan}, {Paczynski}  \& {Piran}}{{Narayan}
  et~al.}{1992}]{Narayan1992}
{Narayan} R.,  {Paczynski} B.,   {Piran} T.,  1992, \mn@doi [\apjl]
  {10.1086/186493}, \href
  {https://ui.adsabs.harvard.edu/abs/1992ApJ...395L..83N} {395, L83}

\bibitem[\protect\citeauthoryear{{Narayan}, {Igumenshchev}  \&
  {Abramowicz}}{{Narayan} et~al.}{2003}]{narayan2003}
{Narayan} R.,  {Igumenshchev} I.~V.,   {Abramowicz} M.~A.,  2003, \mn@doi
  [\pasj] {10.1093/pasj/55.6.L69}, \href
  {https://ui.adsabs.harvard.edu/abs/2003PASJ...55L..69N} {55, L69}

\bibitem[\protect\citeauthoryear{{Paczynski}}{{Paczynski}}{1991}]{Paczynski1991}
{Paczynski} B.,  1991, \actaa, \href
  {http://cdsads.u-strasbg.fr/abs/1991AcA....41..257P} {41, 257}

\bibitem[\protect\citeauthoryear{{Peacock}}{{Peacock}}{1983}]{Peacock1983}
{Peacock} J.~A.,  1983, \mn@doi [\mnras] {10.1093/mnras/202.3.615}, \href
  {https://ui.adsabs.harvard.edu/abs/1983MNRAS.202..615P} {202, 615}

\bibitem[\protect\citeauthoryear{{Perna}, {Armitage}  \& {Zhang}}{{Perna}
  et~al.}{2006}]{Perna2006}
{Perna} R.,  {Armitage} P.~J.,   {Zhang} B.,  2006, \mn@doi [\apjl]
  {10.1086/499775}, \href {http://cdsads.u-strasbg.fr/abs/2006ApJ...636L..29P}
  {636, L29}

\bibitem[\protect\citeauthoryear{{Petropoulou}, {Barniol Duran}  \&
  {Giannios}}{{Petropoulou} et~al.}{2017}]{PBDG2017}
{Petropoulou} M.,  {Barniol Duran} R.,   {Giannios} D.,  2017, \mn@doi [\mnras]
  {10.1093/mnras/stx2151}, \href
  {http://cdsads.u-strasbg.fr/abs/2017MNRAS.472.2722P} {472, 2722}

\bibitem[\protect\citeauthoryear{{Popham}, {Woosley}  \& {Fryer}}{{Popham}
  et~al.}{1999}]{Popham1999}
{Popham} R.,  {Woosley} S.~E.,   {Fryer} C.,  1999, \mn@doi [\apj]
  {10.1086/307259}, \href
  {https://ui.adsabs.harvard.edu/abs/1999ApJ...518..356P} {518, 356}

\bibitem[\protect\citeauthoryear{{Preece}, {Briggs}, {Mallozzi}, {Pendleton},
  {Paciesas}  \& {Band}}{{Preece} et~al.}{2000}]{Preece2000}
{Preece} R.~D.,  {Briggs} M.~S.,  {Mallozzi} R.~S.,  {Pendleton} G.~N.,
  {Paciesas} W.~S.,   {Band} D.~L.,  2000, \mn@doi [\apjs] {10.1086/313289},
  \href {https://ui.adsabs.harvard.edu/abs/2000ApJS..126...19P} {126, 19}

\bibitem[\protect\citeauthoryear{Press \& Teukolsky}{Press \&
  Teukolsky}{1988}]{Press88}
Press W.~H.,  Teukolsky S.~A.,  1988, \mn@doi [Computers in Physics]
  {10.1063/1.4822753}, 2, 74

\bibitem[\protect\citeauthoryear{{Proga} \& {Zhang}}{{Proga} \&
  {Zhang}}{2006}]{Proga2006}
{Proga} D.,  {Zhang} B.,  2006, \mn@doi [\mnras]
  {10.1111/j.1745-3933.2006.00189.x}, \href
  {http://cdsads.u-strasbg.fr/abs/2006MNRAS.370L..61P} {370, L61}

\bibitem[\protect\citeauthoryear{{Qin} et~al.,}{{Qin}
  et~al.}{2013}]{2013ApJ...763...15Q}
{Qin} Y.,  et~al., 2013, \mn@doi [\apj] {10.1088/0004-637X/763/1/15}, \href
  {https://ui.adsabs.harvard.edu/abs/2013ApJ...763...15Q} {763, 15}

\bibitem[\protect\citeauthoryear{{Rowlinson}, {Gompertz}, {Dainotti},
  {O'Brien}, {Wijers}  \& {van der Horst}}{{Rowlinson}
  et~al.}{2014}]{Rowlinson2014}
{Rowlinson} A.,  {Gompertz} B.~P.,  {Dainotti} M.,  {O'Brien} P.~T.,  {Wijers}
  R.~A.~M.~J.,   {van der Horst} A.~J.,  2014, \mn@doi [\mnras]
  {10.1093/mnras/stu1277}, \href
  {https://ui.adsabs.harvard.edu/abs/2014MNRAS.443.1779R} {443, 1779}

\bibitem[\protect\citeauthoryear{{Sobacchi}, {Granot}, {Bromberg}  \&
  {Sormani}}{{Sobacchi} et~al.}{2017}]{sobacchi2017}
{Sobacchi} E.,  {Granot} J.,  {Bromberg} O.,   {Sormani} M.~C.,  2017, \mn@doi
  [\mnras] {10.1093/mnras/stx2083}, \href
  {https://ui.adsabs.harvard.edu/abs/2017MNRAS.472..616S} {472, 616}

\bibitem[\protect\citeauthoryear{{Stanek} et~al.}{{Stanek}
  et~al.}{2003}]{Stanek+03}
{Stanek} K.~Z.,  et~al., 2003, \mn@doi [\apjl] {10.1086/376976}, \href
  {http://adsabs.harvard.edu/abs/2003ApJ...591L..17S} {591, L17}

\bibitem[\protect\citeauthoryear{{Stratta}, {Dainotti}, {Dall'Osso},
  {Hernandez}  \& {De Cesare}}{{Stratta} et~al.}{2018}]{Stratta2018}
{Stratta} G.,  {Dainotti} M.~G.,  {Dall'Osso} S.,  {Hernandez} X.,   {De
  Cesare} G.,  2018, \mn@doi [\apj] {10.3847/1538-4357/aadd8f}, \href
  {https://ui.adsabs.harvard.edu/abs/2018ApJ...869..155S} {869, 155}

\bibitem[\protect\citeauthoryear{{Tang}, {Huang}, {Geng}  \& {Zhang}}{{Tang}
  et~al.}{2019}]{Tang2019}
{Tang} C.-H.,  {Huang} Y.-F.,  {Geng} J.-J.,   {Zhang} Z.-B.,  2019, \mn@doi
  [\apjs] {10.3847/1538-4365/ab4711}, \href
  {https://ui.adsabs.harvard.edu/abs/2019ApJS..245....1T} {245, 1}

\bibitem[\protect\citeauthoryear{{Tchekhovskoy} \& {Giannios}}{{Tchekhovskoy}
  \& {Giannios}}{2015}]{TG15}
{Tchekhovskoy} A.,  {Giannios} D.,  2015, \mn@doi [\mnras]
  {10.1093/mnras/stu2229}, \href
  {https://ui.adsabs.harvard.edu/abs/2015MNRAS.447..327T} {447, 327}

\bibitem[\protect\citeauthoryear{{Tchekhovskoy}, {Narayan}  \&
  {McKinney}}{{Tchekhovskoy} et~al.}{2011}]{2011MNRAS.418L..79T}
{Tchekhovskoy} A.,  {Narayan} R.,   {McKinney} J.~C.,  2011, \mn@doi [\mnras]
  {10.1111/j.1745-3933.2011.01147.x}, \href
  {https://ui.adsabs.harvard.edu/abs/2011MNRAS.418L..79T} {418, L79}

\bibitem[\protect\citeauthoryear{{Tegmark} et~al.,}{{Tegmark}
  et~al.}{2004}]{2004PhRvD..69j3501T}
{Tegmark} M.,  et~al., 2004, \mn@doi [\prd] {10.1103/PhysRevD.69.103501}, \href
  {https://ui.adsabs.harvard.edu/abs/2004PhRvD..69j3501T} {69, 103501}

\bibitem[\protect\citeauthoryear{{Thompson}, {Chang}  \& {Quataert}}{{Thompson}
  et~al.}{2004}]{Thompson2004}
{Thompson} T.~A.,  {Chang} P.,   {Quataert} E.,  2004, \mn@doi [\apj]
  {10.1086/421969}, \href
  {https://ui.adsabs.harvard.edu/abs/2004ApJ...611..380T} {611, 380}

\bibitem[\protect\citeauthoryear{{Troja} et~al.,}{{Troja}
  et~al.}{2007}]{Troja2007}
{Troja} E.,  et~al., 2007, \mn@doi [\apj] {10.1086/519450}, \href
  {http://adsabs.harvard.edu/abs/2007ApJ...665..599T} {665, 599}

\bibitem[\protect\citeauthoryear{{Usov}}{{Usov}}{1992}]{Usov1992}
{Usov} V.~V.,  1992, \mn@doi [\nat] {10.1038/357472a0}, \href
  {http://cdsads.u-strasbg.fr/abs/1992Natur.357..472U} {357, 472}

\bibitem[\protect\citeauthoryear{{Wanderman} \& {Piran}}{{Wanderman} \&
  {Piran}}{2010}]{wanderman_piran2010}
{Wanderman} D.,  {Piran} T.,  2010, \mn@doi [\mnras]
  {10.1111/j.1365-2966.2010.16787.x}, \href
  {http://adsabs.harvard.edu/abs/2010MNRAS.406.1944W} {406, 1944}

\bibitem[\protect\citeauthoryear{{Woosley}}{{Woosley}}{1993}]{Woosley93}
{Woosley} S.~E.,  1993, \mn@doi [\apj] {10.1086/172359}, \href
  {http://adsabs.harvard.edu/abs/1993ApJ...405..273W} {405, 273}

\bibitem[\protect\citeauthoryear{{Woosley} \& {Bloom}}{{Woosley} \&
  {Bloom}}{2006}]{Woosley&Bloom06}
{Woosley} S.~E.,  {Bloom} J.~S.,  2006, \mn@doi [\araa]
  {10.1146/annurev.astro.43.072103.150558}, \href
  {http://adsabs.harvard.edu/abs/2006ARA%26A..44..507W} {44, 507}

\bibitem[\protect\citeauthoryear{Zhang, Woosley  \& MacFadyen}{Zhang
  et~al.}{2003}]{zhang2003}
Zhang W.,  Woosley S.~E.,   MacFadyen A.~I.,  2003, The Astrophysical Journal,
  586, 356

\bibitem[\protect\citeauthoryear{{Zhang}, {Fan}, {Dyks}, {Kobayashi},
  {M{\'e}sz{\'a}ros}, {Burrows}, {Nousek}  \& {Gehrels}}{{Zhang}
  et~al.}{2006}]{Zhang2006}
{Zhang} B.,  {Fan} Y.~Z.,  {Dyks} J.,  {Kobayashi} S.,  {M{\'e}sz{\'a}ros} P.,
  {Burrows} D.~N.,  {Nousek} J.~A.,   {Gehrels} N.,  2006, \mn@doi [\apj]
  {10.1086/500723}, \href {http://adsabs.harvard.edu/abs/2006ApJ...642..354Z}
  {642, 354}

\bibitem[\protect\citeauthoryear{{Zhang}, {Zhang}, {Murase}, {Connaughton}  \&
  {Briggs}}{{Zhang} et~al.}{2014}]{2014ApJ...787...66Z}
{Zhang} B.-B.,  {Zhang} B.,  {Murase} K.,  {Connaughton} V.,   {Briggs} M.~S.,
  2014, \mn@doi [\apj] {10.1088/0004-637X/787/1/66}, \href
  {https://ui.adsabs.harvard.edu/abs/2014ApJ...787...66Z} {787, 66}

\makeatother
\end{thebibliography}






\bsp	
\label{lastpage}
\end{document}